# Organic Molecular Thin Films for Nanoscale Information Memory Applications

**J. C. Li***

*Vacuum and fluid engineering research center, School of mechanical engineering and Automation, Northeastern University, Shenyang 110004, China*

## Table of Contents

*Corresponding author: jianchang_li@hotmail.com

## Table of Abbreviations

ACE, acetone
AcS, thiolacetyl
AFM, atomic force microscopy
Ag, silver
Au, gold
CAFM, contacting atomic force microscopy
Cu, copper
CMOS, complementary metal-oxide-semiconductor transistor
CT, charge transfer
DI, de-ionized
DFT, density functional theory
EA, electron affinity
Et, Ethyl
EtOH, ethanol
GaIn, gallium indium
H, hour
HAc, acetic acid
HF, hydrofluoric acid or Hartree-Fock
Hg, mercury
$H_2O_2$, hydrogen peroxide
HOMO, highest occupied molecular orbital
HOPG, highly oriented pyrolitic graphite
$H_2SO_4$, sulfuric acid
IE, ionization energy
IP, ionization potential
IPES, inverse photoemission spectroscopy
ITO, indium tin oxide
I-V, current-voltage
LB, langmuir-Blodgett
LED, light emitting diode
LUMO, lowest unoccupied molecular orbital
Me, Methyl
ML, milliliter
mM, millimole
MOM, metal-organic-metal
NN, nano-Newton
OPV, oligophenylenevinylene
PMMA, polymethylmethacrylate
RCA, radio corporation of america
Redox, reduction oxidation
Rpm, rotation per minute
SAM, self-assembly monolayer
SAMs, self-assembly monolayers
SPM, scanning probe microscopy
STM, scanning tunneling microscopy
STS, scanning tunneling spectroscopy
THF, tetrahydrofuran
UPS, ultraviolet photoemission spectroscopy
V, voltage
$V_t$, turn-on voltage
W, watt

## Table of Symbols

$J$, current density
$F$, electrical field
$\phi_B$, the zero-field injection barrier
$q$, is elementary charge
$m^*$ effective electron mass
$k_B$, Boltzmann's constant
$h$, Planck's constant
$T$, temperature
$\varepsilon$, relative dielectric constant
$\varepsilon_0$, vacuum permittivity
$e$, the charge of an electron
$\hbar$, Plank's constant divided by $2\pi$
$d$, the tunneling distance (or wire length in the case of molecular junctions)
$\phi$, the barrier height
$V$, the bias voltage applied between the electrodes
$m$, the mass of an electron
$C$, the proportionality constant
$\alpha$, a unitless adjustable parameter used in fitting
$I$, current
$\beta$, decay constant

## Abstract

According to Moore's law, the silicon semiconductor transistor based information system is facing its physical limitations due to fluctuations of random charge and leakage current. Molecular electronics is becoming increasingly attractive owing to the advantages of easy molecular structure variability, flexibility, low-cost and compatibility with bioelectronics. In this handbook chapter, we reviewed the recent research progress of molecular electronics, especially the studies on nanoscale information memories, from the viewpoints of structure-property relationship. Two kinds of molecular systems including redox dendrimeric thin films and self-assembled molecular monolayers are discussed in detail. The investigation and application of other molecular thin films such as polymers, charge transfer salts and Langmuir-Blodgett layers are briefly introduced. We suggest that two promising molecular systems have the most potentials for using as building blocks in nanoscale information storage. One is single-molecule-based memory device with sub-10 nm characteristics built on self-assembled monolayer. Multimode information storage is the other powerful way to make breakthrough in the challenging area of nanoscale data storage. This relies on further experimental and theoretical advances. Moreover, a big foreseeable obstacle is how to bridge the big gap between such novel system and the current bit world.

## 7.1 Introduction

Silicon-based semiconductor industry is very successful in the past decades, and has led us to an information age. According to Moore's law, the semiconductor device density in a chip has to be doubled every one and half year [1]. To date, the semiconductor devices have been miniaturized to sub-50 nm dimensions, where several factors are limiting the bit density increase including lithography resolution, electromigration, and capacitance. With the onslaught of digital age, the demands of ultrahigh data density have fuelled the need for information memory media with sub-10 nm scale that falls to the dimensions of a single molecule [2,3]. However, there are some crucial effects emerging from such low dimension devices: 1) when too few atoms are used in a wire they move in response to currents; 2) when too few electrons are used, the fluctuation in their number becomes significant in the device performance; 3) the electron noise in a nanoscale world may come from thermodynamic fluctuations, defect scattering, and finite-size statistics.

To develop ultra-high density information storage systems, some other critical factors have to be taken into account too: density, cost, writing/erasing time, reversibility, power consumption and durability. Now, researchers are trying to continue the current trend of device miniaturization to fulfill the Moore's law requirement. Such an effort calls for the development of new techniques, materials and theories for nanoscale information storage. An important technique of scanning tunneling microscopy (STM) was invented in 1980s. The development of various scanning probe microscope (SPM) techniques has brought with the possibilities to construct and further characterize molecular junctions with different conformation, substituent and molecule number within the junction. From then on, in the past 20 years it has been seen that there are the rapid and significant advances of both basic and applied research in the field of molecular electronics. As a bottom-up approach in molecular electronics, we have to exploit the possibility of using organic functional molecules as active information storage media, which includes molecular design, synthesis, active media growth, device fabrication and structure-property characterizations [4]. This has opened up a new research area in the field of molecular electronics.

Since Avriam and Ratner proposed the donor-sigma-acceptor model of single molecular diodes in 1970s, molecular materials including molecular clusters and thin layers have been being intensively studied owing to their unique optical/electrical properties and potential applications in the field of new generation of information technology. The molecular switching performance, especially optoelectronic switching, of these nanostructures are the very important functional features for the usage of information memories. To understand the switching mechanisms, one has to construct a molecular device, usually in the form of metal-organic-metal (MOM) junction. The current-voltage (I-V) characteristics of the molecular junction can be then investigated under various controlled conditions (e.g. temperature, gas and/or solvents, and optical illumination etc). The correlation between these researching contents is schematically shown in Figure 1.

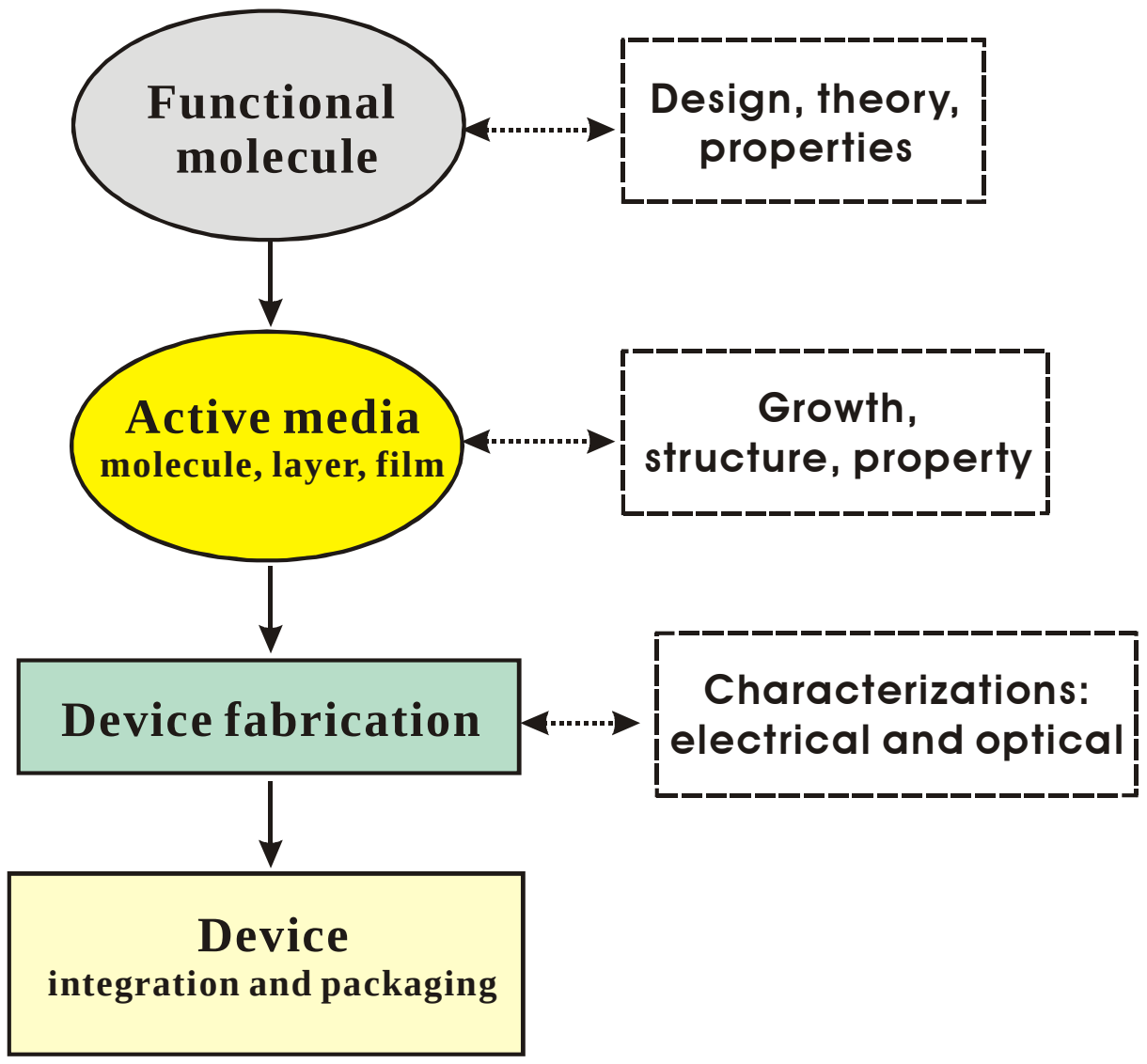

**Figure 1.** Schematic correlation between the main researching contents of molecular electronics.

In this chapter, we have reviewed functional molecules for application in the field of information storage, i.e. with molecular switching performance, from the aspects of chemical conformation/substituent, thin film growth and structure-property relationships. Two kinds of molecular materials are discussed in detail, which include redox dendrimeric materials and self-assembly monolayer (SAM). In addition, the fabrication and growth of polymer, LB, and charge transfer salts-based thin films/devices are briefly discussed too. For easy reference of this handbook chapter, most of the data is presented in the forms of table and figure.

The organization of this chapter is as follows. First, we give a brief presentation of the techniques for organic active layer growth, device fabrication and molecular junction switching mechanism. Especially, more emphasis is given to the fabrication of self-assembled monolayers and junctions. Second, we will discuss the redox dendrimeric materials based molecular memories with highlight on the structure-property relationship and ultra-high density charge storage. Then as the focus of this chapter, we describe the experimental studies on self-assembled monolayer-based molecular memories from the aspects of molecular structure-property relationship, device fabrication and characterizations. Finally, we will briefly introduce the usage of polymers, LB layers and charge transfer salts as functional media for molecular memories.

## 7.2 Active Layer Growth, Device Fabrication and Switching Mechanism

This section generally introduces the techniques and knowledge about thin film deposition, growth of monolayer, fabrication and characterization of molecular junctions. The switching mechanism for usual molecular electronic devices is also briefly presented, which will be further discussed in the later sections.

### 7.2.1 Active Layer Growth

There are several types of molecular active media for potential application in information memory devices, which include thin films, self-assembled monolayers, Langmuir-Blodgett (LB) layers, molecular patterns and single molecules. In this section, the techniques for active layer fabrication are briefly presented in the sequence of the frequency used and significance.

#### *7.2.1.1 Vacuum Thermal Vapor Deposition*

Vacuum thermal evaporation is a traditional method widely used for fabrication of various thin films including metal electrodes and organic layers. It has the advantages of clean and controlled environment (excluding the factors of water and oxygen), known thickness, multi-sources and in-situ fabrication and characterization etc. Figure 2 shows a schematic vacuum system that we use for the in-situ fabrication and characterization of molecular electronic devices. The chamber is pumped by a combined vacuum station consisting of a turbo pump and a mechanical pump. There are two separate evaporating stations within the chamber and each one has two evaporation sources. The substrate is sheltered during the exchange of the sources when a layer deposition is finished. The substrate can be cooled down to liquid nitrogen temperature during the metal electrode deposition so as to protect the organic layer. The film thickness, substrate temperature and optical illumination will be continually monitored and controlled during the film fabrication and device characterization processes by using a computer through a data board. The application of such systems is further discussed in the section of self-assembled monolayers.

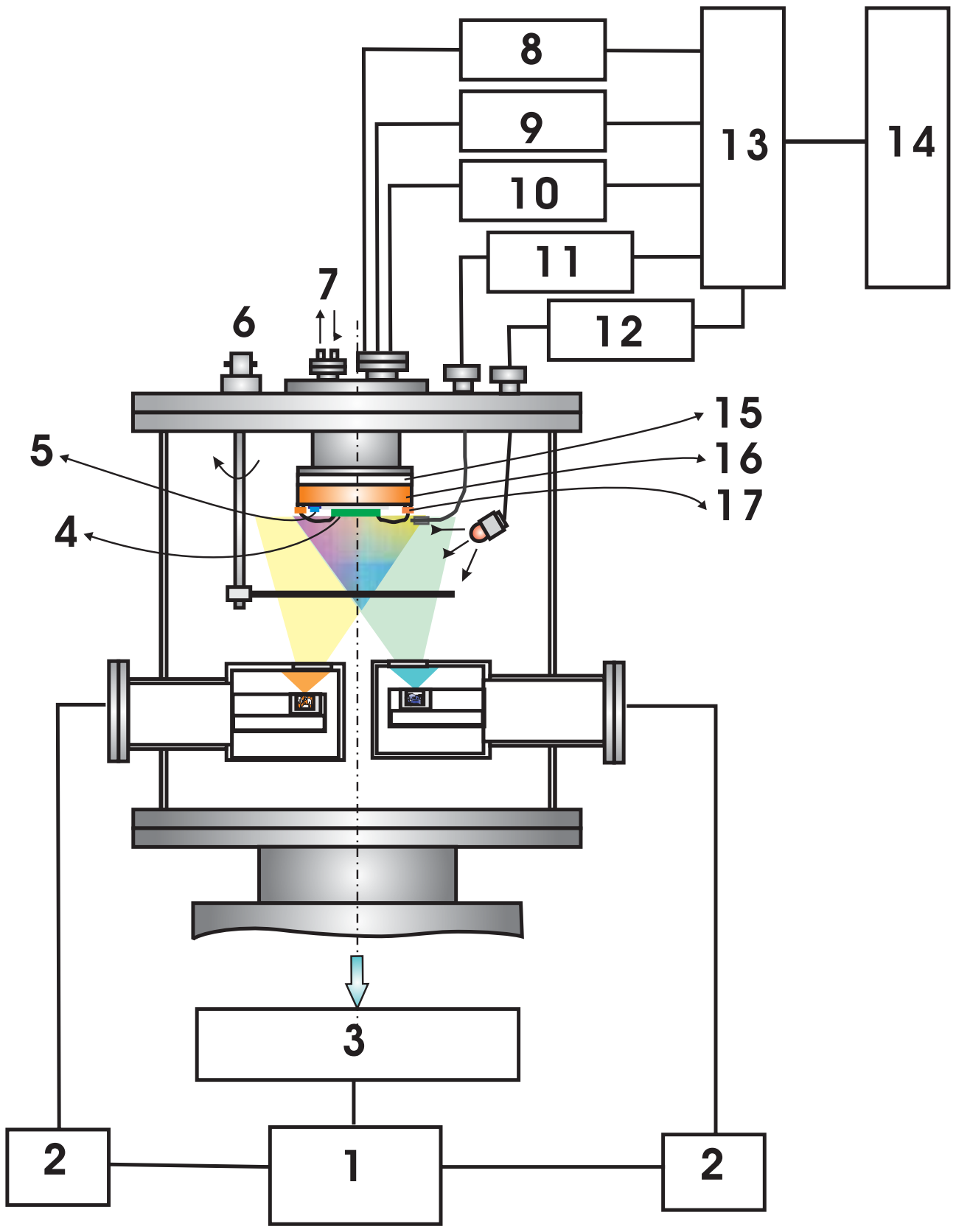

**Figure 2.** Schematic vacuum thermal vapor deposition system with in-situ vary-temperature current-voltage characterizations. Here, 1– Power, 2–evaporation crucible heating controllers, 3– to ultra-high vacuum pumps, 4– sample, 5– temperature sensor, 6– shutter, 7- liquid nitrogen feedthrough, 8– current/voltage measurement, 9– temperature detector, 10– substrate heating controller, 11– film thickness monitor, 12– LED light, 13– data processing board, 14–computer, 15– thermal insulator, 16– substrate holder, 17– electrical bridging connector.

### *7.2.1.2 Self-Assembly Monolayer*

Self-assembly monolayer (SAM) method is a promising approach for the mass production of organic electronic devices with nanoscale dimension, high reproducibility and low-cost advantages, which is preferable in the next generation of ultra-high density information memory applications [5-8]. There are usually two kinds of approaches to make self-assembled monolayers in the field of molecular electronics. One is based on thiolated molecular materials that can be easily self-assembled onto fresh deposited gold surface. The other method is to make monolayers on silicon surface with or without thermal dioxide layer through organosilane chemical process. We first introduce some necessary knowledge about organic solvents that are used in the preparation of molecular monolayers. The common solvents include methanol, ethanol (EtOH), propanol, butanol, acetone (ACE), tetrahydrofuran (THF), acetic acid (HAc) and chloroform etc. Their formula and physical properties are given in Table 1. *Caution! The handle of hydrofluoric acid (HF) and sulfuric acid need to be very careful with full personal protection and in well-ventilated chemical hood.*

**Table 1.** Common organic solvents used for molecular monolayer and/or substrate treatments.

| **Solvent (abbr.)** | **Chemical formula** | **Boling point (°C)** | **Solubility in 25 °C water** | **Toxic, corrosive and dangerous scale** |
|---|---|---|---|---|
| Methanol | $CH_3OH$ | 65 | Completely | ××× |
| Ethanol (EtOH) | $CH_3$ $CH_2OH$ | 78 | Completely | Low |
| Propanol | $CH_3$ $CH_2CH_2OH$ | 97 | Completely | ×× |
| Butanol | $CH_3$ $CH_2CH_2CH_2OH$ | 118 | 0.08 g/cm$^3$ | ××× |
| Acetone (ACE) | $CH_3$ $COCH_3$ | 58 | Completely | Low |
| Acetic acid | $CH_3$ COOH | 118 | Infinite | ××× |
| *Chloroform | $CHCl_3$ | 61.7 | Not | ×××× |
| Tetrahydrofuran (THF) | O $CH_2CH_2CH_2$ | 65.4 | Not | Low |
| *Sulfuric acid | $H_2SO_4$ | 338 | Completely | ×××× |
| *Hydrogen peroxide | $H_2O_2$ | 152.1 | Completely | ×××× |
| *Hydrofluoric acid | HF | 112.2 | 1.14 g/ cm$^3$ | ××××× (Extremely hazardous) |

*Note: these solvents need to be handled carefully with full protections from any direct contact with skin, eye or inhalation. ××××× and ××× indicate the levels of hazard.

## A. Thiolated material-based monolayers

Table 2 lists out the experimental procedure to make n-alkanethiol SAMs onto gold substrate surface. The key points are substrate clean (if not freshly deposited) and concentration of target molecular solution. To make high quality monolayer, the gold film is always deposited just before the growth of SAMs. One mille-mole molecular solution is usually an appropriate concentration for the fabrication of n-alkanethiol SAMs. Figure 3 shows two schematic molecular junctions made from alkanethiol and alkanedithiol SAM between gold electrodes. It is obvious that the top molecule/electrode interfaces for the two kinds of SAMs are different, which will inevitably affect the device performance.

The experimental conditions and growth procedure for conjugated molecular SAMs are different due to their relatively high rigidity and non-flexible natures comparing to that of the alkanethiol counterparts. In this case, it will be a good idea to make conjugated molecular SAMs with densely packing characteristics out of solution mixed with a few percent of alkanethiols. This has been named as monolayer matrix technique that will be further discussed in the later corresponding section of this chapter.

**Table 2.** Procedure to make thiolated molecular self-assembly monolayer onto Au surface. Here 1-decanethiol (i.e. $C_{10}SH$) is used as a typical example.

| **Entry** | **Action** | **Note** |
|---|---|---|
| 1 | Prepare 1 mM target molecular solution; usually | Make sure the gold substrate and container |
| | pure ethanol is used as the solvent. | have been previously cleaned as the process given before. |
| 2 | Immerse the Au substrate into the molecular solution and de-oxygen by using nitrogen gas. | Keep it in dark for 18 to 24 h to allow the SAMs growth. |
| 3 | Take out the sample and rinse with the sequential flow of pure ethanol, DI water and pure ethanol, respectively. | About 60 to 90 seconds for each steps. |
| 4 | Dry with nitrogen gun | 60 s |
| 5 | Use the sample immediately. | Otherwise, keep it in pure ethanol. |

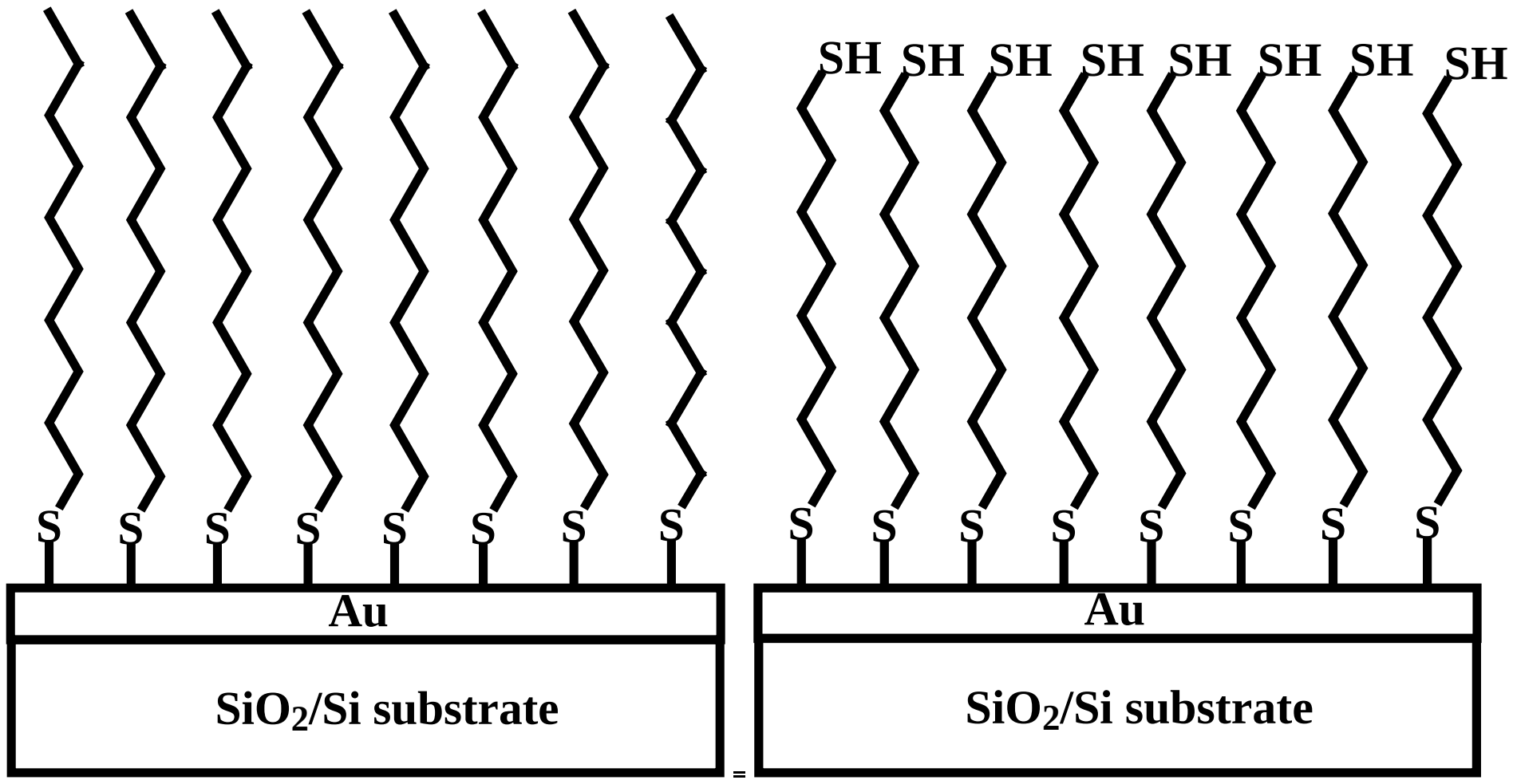

**Figure 3.** Schematic self-assembly monolayer of (a) 1-decanethiol and (b) 1-nonanedithiol sandwiched between Au electrodes.

### B. Organosilane-based molecular monolayers

To make silane monolayer, the silicon wafer need to be carefully cleaned and pre-treated with flat dry thermal dioxide surface. The wafer Radio Corporation of America (RCA) clean is a standard set of wafer cleaning steps which needs to be performed before high temperature processing steps (e.g. oxidation) of silicon wafers in semiconductor manufacturing. Werner Kern developed this basic procedure in 1965 while working for the RCA. It involves the following steps: 1) Removal of the organic contaminants (i.e. organic clean); 2) Removal of thin oxide layer (i.e. oxide strip); 3) Removal of ionic contamination (i.e. ionic clean). The first step is performed with a 1:1:5 solution of $NH_4OH$ + $H_2O_2$ + DI $H_2O$ at 75 to 80 ℃ for 15 minutes. This treatments results in the formation of about 1 nm thin silicon dioxide layer on the silicon surface, along with a certain degree of metallic contamination (especially Iron) that shall be removed in subsequent steps. The second step is a short immersion of the wafer in a 1:50 solution of HF : DI $H_2O$ at 25 ℃ for no more than 10 seconds, in order to remove the thin oxide layer and some fraction of ionic contaminants. The third step is performed with a 1: 1: 6 solution of HCl: $H_2O_2$ : DI $H_2O$ at 75 to 80 ℃ for about 10 minutes. This procedure can effectively removes the remaining traces of metallic ionic contaminants from the silicon wafer surface. The experimental details of RCA clean are summarized in the following table 3.

**Table 3.** Procedure for silicon wafer RCA clean.

| Step | Solution and ratio | Temperature (˚C) | Time | Purpose |
|---|---|---|---|---|
| 1 | $NH_4OH$ : $H_2O_2$ : DI $H_2O$ = 1:1:5 | 75 ~ 80 | ~ 15 min | Organic clean |
| 2 | HF : DI $H_2O$ = 1:50 | 25 | ~ 10 s | Remove oxide |
| 3 | HCl + $H_2O_2$ +DI $H_2O$ = 1:1:6 | 75 ~ 80 | ~ 10 min | Clean metallic ionic |

After wafer RCA clean, the substrate is now ready for organosilane monolayer growth. Table 4 describes a regular experimental procedure to make tricholrosilane monolayer onto Si(100) substrate surface [9-12]. As the thickness for such a monolayer is just a few nanometers, the assembly side of the substrate should be pre-polished and atomic flat. One millimolar molecualr solution usually yields a good monolayer result with growth time of 12 to 24 h under ambient conditions. To avoid overlayers, the sample has to be intensively rinsed with the same solvent as used for the solution and then isopropyl alcoho and di-ionized water. This process is shematically shown in Figure 4.

**Table 4.** Experimental procedure to make organosilane monolayer (e.g. trichlorosilane) on silicon dioxide substrate.

| **Step** | **Action** | **Note** |
|---|---|---|
| 1 | Silicon wafer RCA clean | Refer to the RCA clean table |
| 2 | Growth of dry thermal dioxides | 800 °C, 30 min in $N_2$ gas to grow several nm silicon dioxide layer |
| 3 | Immerse immediately the treated substrate into target molecular solution and de-oxygen with nitrogen gas flow. | Dip in 1 mM molecular solution for 12 to 24 h |
| 4 | Rinse the sample thoroughly to avoid overlayers | Using solvents chloroform, Isopropyl alcohol and DI water, respectively (few minutes for each step). |
| 5 | Dry with nitrogen gun | 1 min |
| 6 | Annealing in vacuum | 150 °C for 10 to 15 min to improve the monolayers quality and get rid of water and solvent. |
| 7 | A substituted treatment is to bake the sample under nitrogen environment | 75 to 80 °C for 10 min, keep the sample covered (e.g. with aluminum foil) during this process. |

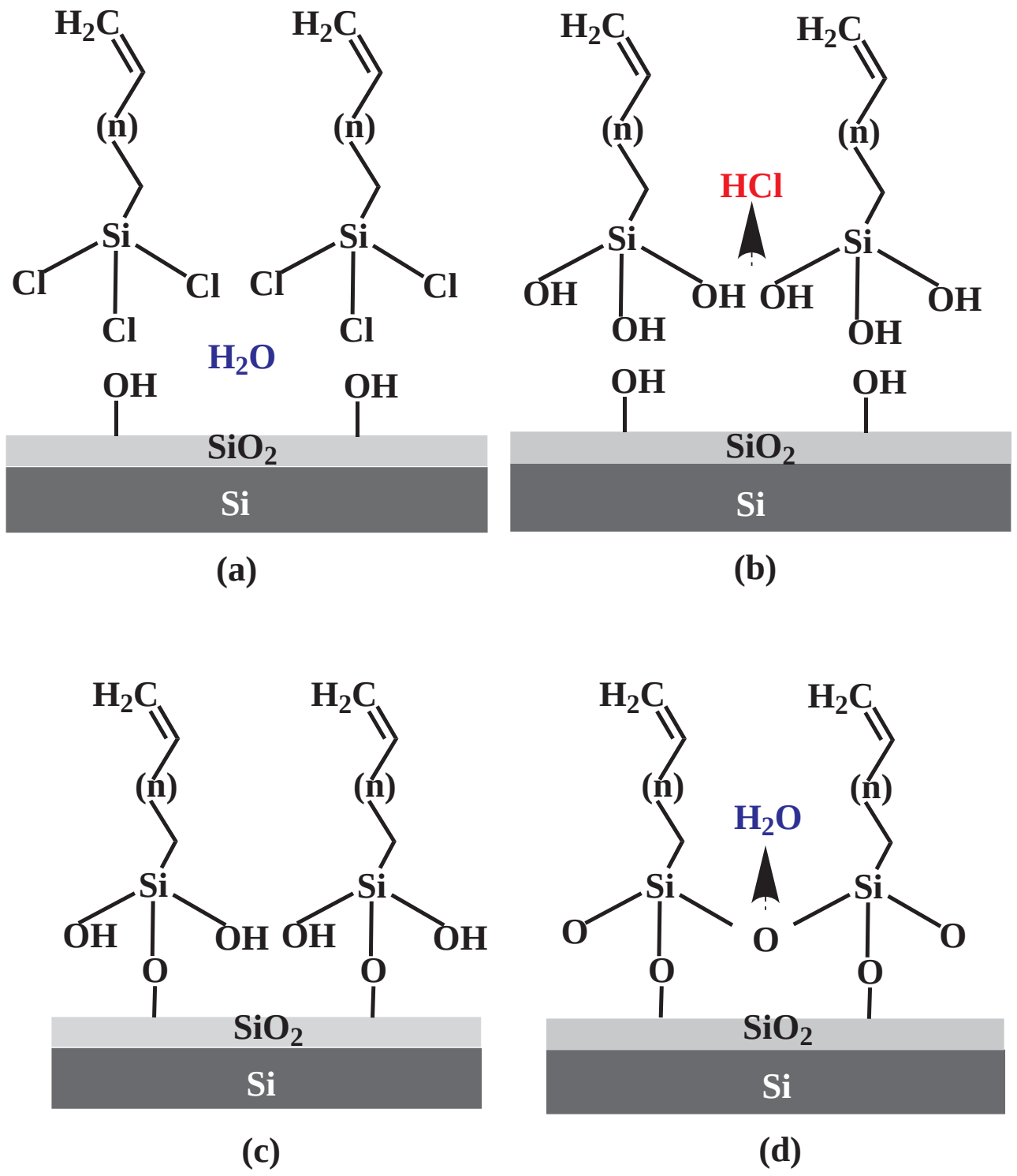

**Figure 4.** Schematic self-assembly process for trichlorosilane monolayer formation on silicon dioxide surface. A very thin water layer is a necessary factor in this chemical process. [Adapted with permission from Ref. 9. Copyright (1997), Elsevier B.V.]

### *7.2.1.3 Langmuir-Blodgett Monolayers*

To make Langmuir-Blodgett monolayer, minute droplets of pre-prepared target molecular solution are carefully cast on the pure water surface in the LB trough at intervals of tens of seconds. As shown in figure 5, the hydrophobic chain-ended molecules remained on the surface of water after evaporation of the solvent. The molecules are then compressed by moving the barriers at a speed of 3 to 6 mm per minute. The surface pressure isotherm has to be recorded during the compression. The target molecules are transferred to the substrate by retracting the substrate, which is pre-immersed in the pure water before the molecular solution is spread in the trough. The retracting speed is kept at about 0.3 to 0.7 mm per minute. The temperature of the pure water in the trough is kept at about 20 to 25 °C.

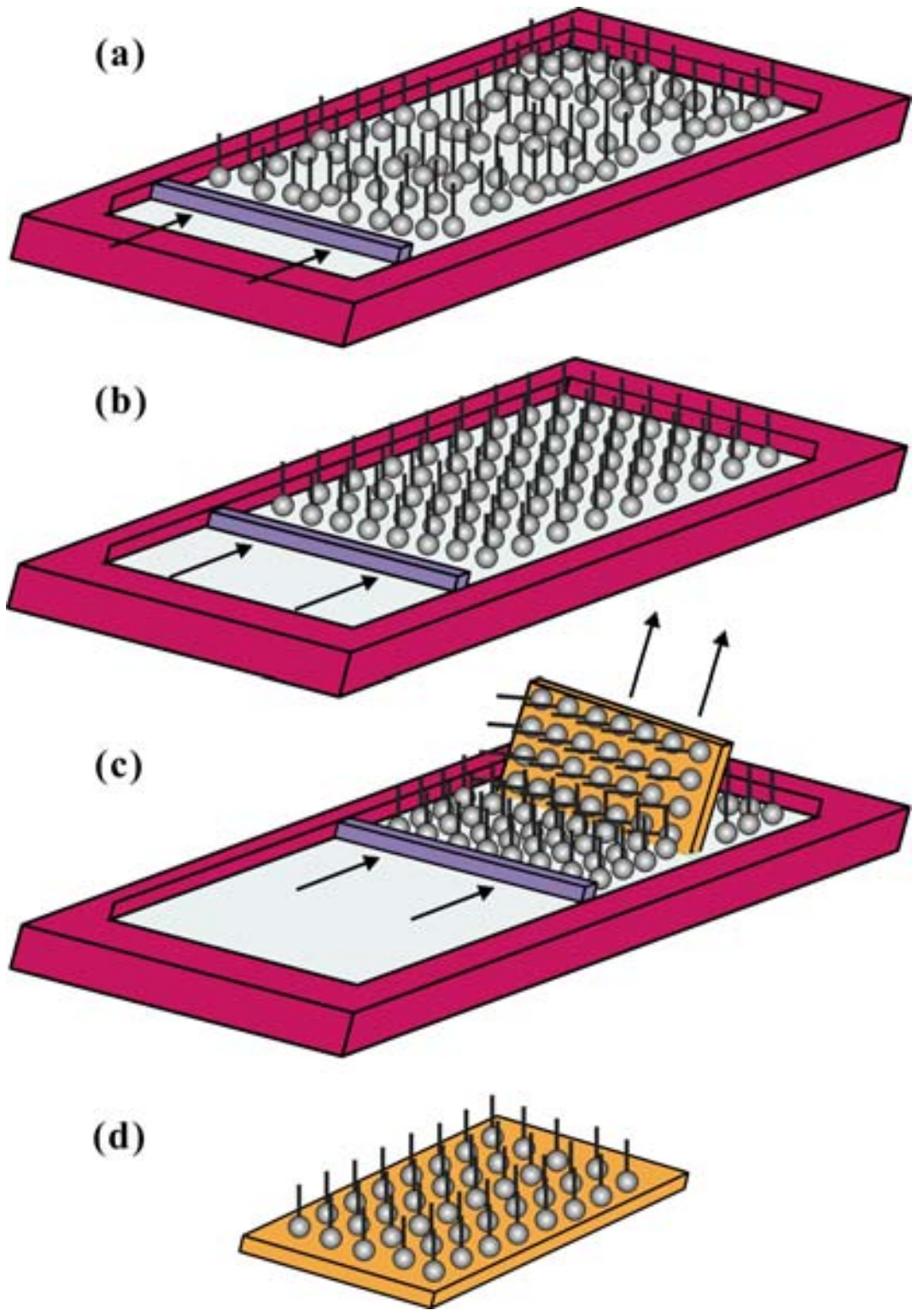

**Figure 5.** Schematic process for LB monolayer fabrication. (a) droplet of molecular solution is spread on the water surface in the trough. (b) The molecules are compressed by slowly moving the barrier bar. (c) The pre-dipped substrate is pull out of the water with the molecules transferred onto its surface. (d) The as-fabricated sample with a single LB layer.

### *7.2.1.4 Vacuum Spray Deposition*

Vacuum spray deposition is a new technique for fabrication of high molecular-weight polymer thin films with quality better than that of the other approaches. As shown in figure 6, to make a spray film, the polymer pre-dissolved in a suitable solvent is firstly fed into a glass-bottle with two openings. The concentration of the dye can be varied according to experimental requirements. The chamber is vacuumed by using a rotary pump through a liquid nitrogen trap, whose pressure is monitored with a vacuum gauge. A valve separates the spray nozzle specially designed with a small mouth (10 to 20 μm in diameter) from the high performance liquid chromatography pump in order to control the spray process. A flow counter is used to monitor the spray rate of the polymer solution through the nozzle. The substrate is heated to 137-157 ℃ before the spray process. When the polymer solution is introduced into the deposition room through the nozzle, it will be aspirated into extremely fine aerosol consisting of numerous of fine droplets with diameters less than 100 nm. The solvent can be evaporated very fast due to the factors of vacuum and high substrate temperature. The polymer is thus adsorbed onto the substrate surface forming a smooth thin film.

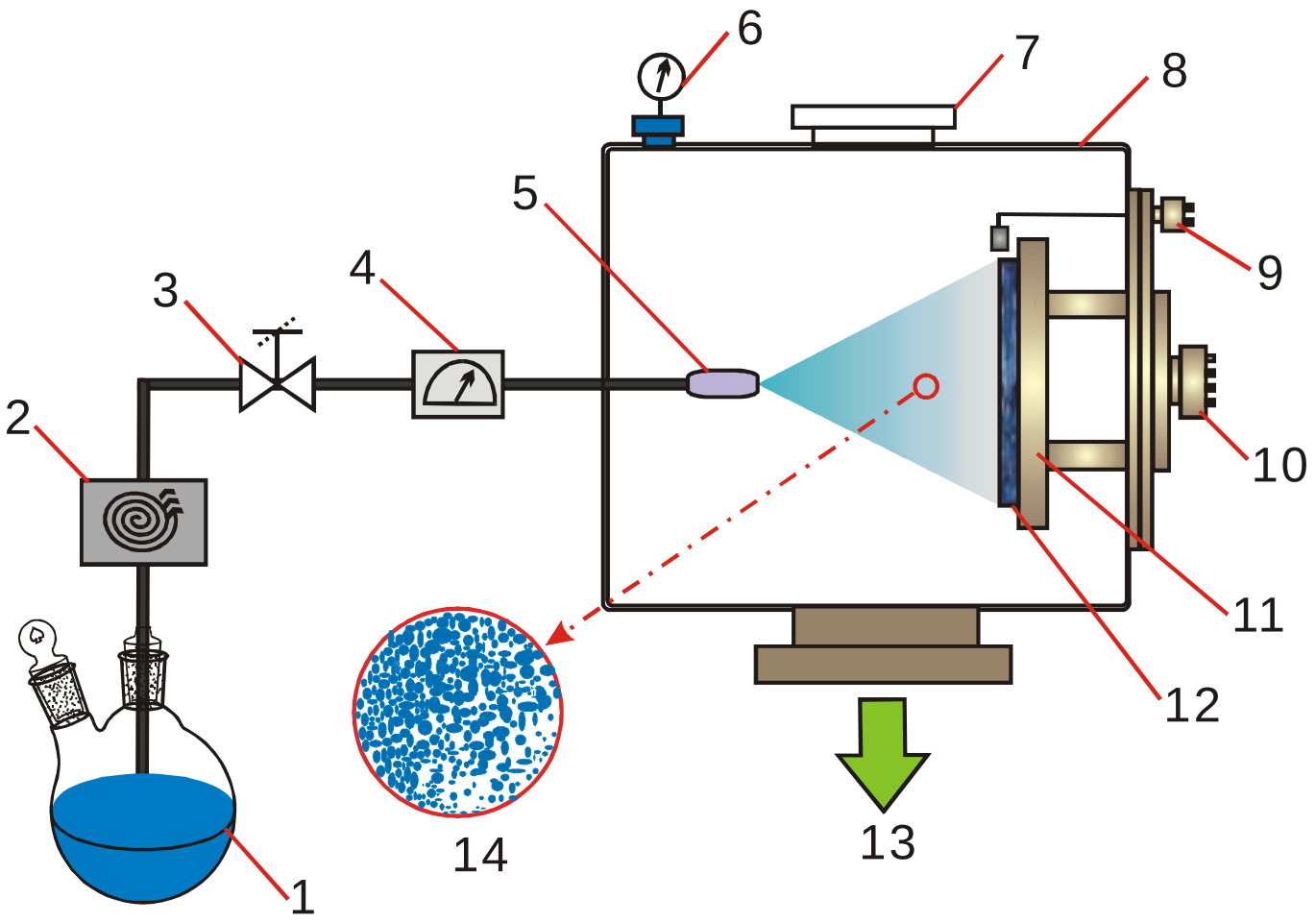

**Figure 6.** Schematic diagram of vacuum spray deposition system. 1–Polymer solution, 2–Compress pump, 3–Valve, 4–Flow meter, 5–Nozzle, 6–Pressure gauge, 7–Observation window, 8–Vacuum chamber, 9–Film thickness monitor, 10–Substrate heating electrical feedthrough, 11–Substrate holder, 12–Sample, 13–To high vacuum pump system and 14–Zoom of the spray.

#### *7.2.1.5 Spin Coating and Dip Coating*

Spin coating and dip coating are some other fabrication methods involved the usage of solution. The quality and thickness of the molecular thin film are mainly controlled through the spinning/dipping speed and molecular concentration. In addition, the baking (or annealing) of the sample under a vacuum condition is also a necessary procedure to get rid of the residue solvent. Like that of vacuum spray, these two approaches are suitable for thin film fabrication from high molecular weight organic materials with high melting points, which tend to be thermally decomposed in traditional vacuum vapor deposition process. Unlike that of the vacuum spray method, they are easy and cheap way with no sophisticated growth systems needed.

### 7.2.2 Device Fabrication

Generally, two terminal molecular junctions are the common type of devices used in the investigation of various molecular electronic systems. A molecular junction consists of a molecular active layer connected with two electrodes that can be made from metal or semiconductor materials. To make a molecular device, firstly we need to prepare a clean substrate before beginning further experimental procedures such as electrode deposition, active layer growth, junction fabrication and characterization.

#### *7.2.2.1 Preparations of Substrate and Electrode*

The regular conductive substrates used in the researches of molecular electronics include at least doped silicon wafer, highly oriented pyrolitic graphite (HOPG), Indium tin oxide (ITO), and metal film (such as Au, Ag, Al and Cu) deposited onto mica, glass or silicon dioxide. In this section, we simply introduce how to prepare various substrates for molecular junction fabrication. Table 5 gives another wafer cleaning procedure besides the standard RCA method, where a step involved with HF acid is used to etch the native oxide layer off the silicon wafer.

**Table 5.** Silicon wafer clean procedure for thermal oxidation or thin film deposition.

| Step | Action | Time (s) |
|---|---|---|
| 1 | Place the wafer in a sample holder | N/A |
| 2 | Place the holder in a 400 ml beaker with acetone and then set it in | 180 |
| | ultrasonic bath | |
| 3 | Place the holder in a 400 ml beaker of isopropyl alcohol | N/A |
| 4 | Set the beaker in the ultrasonic bath | 180 |
| 5 | Rinse with DI water and blow dry with nitrogen | N/A |
| 6 | Clean with *Piranha etch (Caution! Keep away from this explosive and corrosive solution)* | N/A |
| 7 | Cover wafer and take to HF acid bench (*Caution! wear safety apron, gloves and face shield)* | N/A |
| 8 | Place sample holder in 400 ml plastic beaker filled with diluted 10:1 solution DI water : HF acid *(Caution! Extremely hazardous, keep away from any skin contact)* | 60 |
| 9 | Place sample holder in DI water cascade and rinse at each stage | 120 |
| 10 | Dry with nitrogen gun and cover wafer | N/A |
| 11 | Immediately follow with the next experimental procedure | N/A |

Mica and HOPG substrates are normally prepared through fresh cleaving just before the relevant experiment like metal electrode deposition. Sometimes glass slide is used, which can be cleaned following the procedures: 1) Ultrasonically wash with detergent plus de-ionized (DI) water for 3 min, 2) Rinse with DI water flow. 3) Clean with Piranha etch, i.e. dipping into hot $H_2SO_4$ and DI $H_2O_2$ mixed solution (volume ratio 4:1) for about 10 minutes followed by rinsing with water, acetone, and ethanol, respectively.

ITO glass substrates can be cleaned sequentially in ultrasonic baths of acetone and 2-propanol, before oxygen plasma etching under a pressure of 100 Torr at the power of 100 W for about 4 to 6 minutes.

### *7.2.2.2 Molecular Materials*

To realize organic molecule-based information storage, chemists have designed various unique molecular materials with specific opto/electro functional groups and units. The first kind of prime molecules is charge transfer materials with electron donor-acceptor groups, which can be bridged with each other by either a $\sigma$ or a $\pi$ linker. In some cases, co-deposition is used to fabricate complex thin film out of two different molecular materials with electron donor and acceptor groups, respectively. Since there are many intensive review articles and books on charge transfer materials and devices, it is not included in this chapter.

Dendrimeric molecules with core-shell structure are the second kind of important active materials for application as information memory media. The redox gradient nature of dendrimers enables them preferable for single-molecule charge storage application [13]. Self-assembled monolayers based on thiolated and organosilane molecular materials are the third kind of molecular materials with fantastic properties, which make them as an ideal active media for nanoscale information memory device [14]. With using SAMs techniques, scientist can easily make high quality monolayers and even nanoscale patterns on various metal and semiconductor surfaces [6]. Other molecular materials with potential for information memory applications are polymers and LB layers, which will be discussed in the later section.

The molecular structure including functional groups and conformation can greatly affect the device performance and application. Figure 7 shows the building units used in molecular design and construction, while Table 6 lists out the functional substituents usually encountered in molecular electronics research. These knowledge will be a necessary preparation for later sections when we discuss the molecular and/or device structure-property relationships.

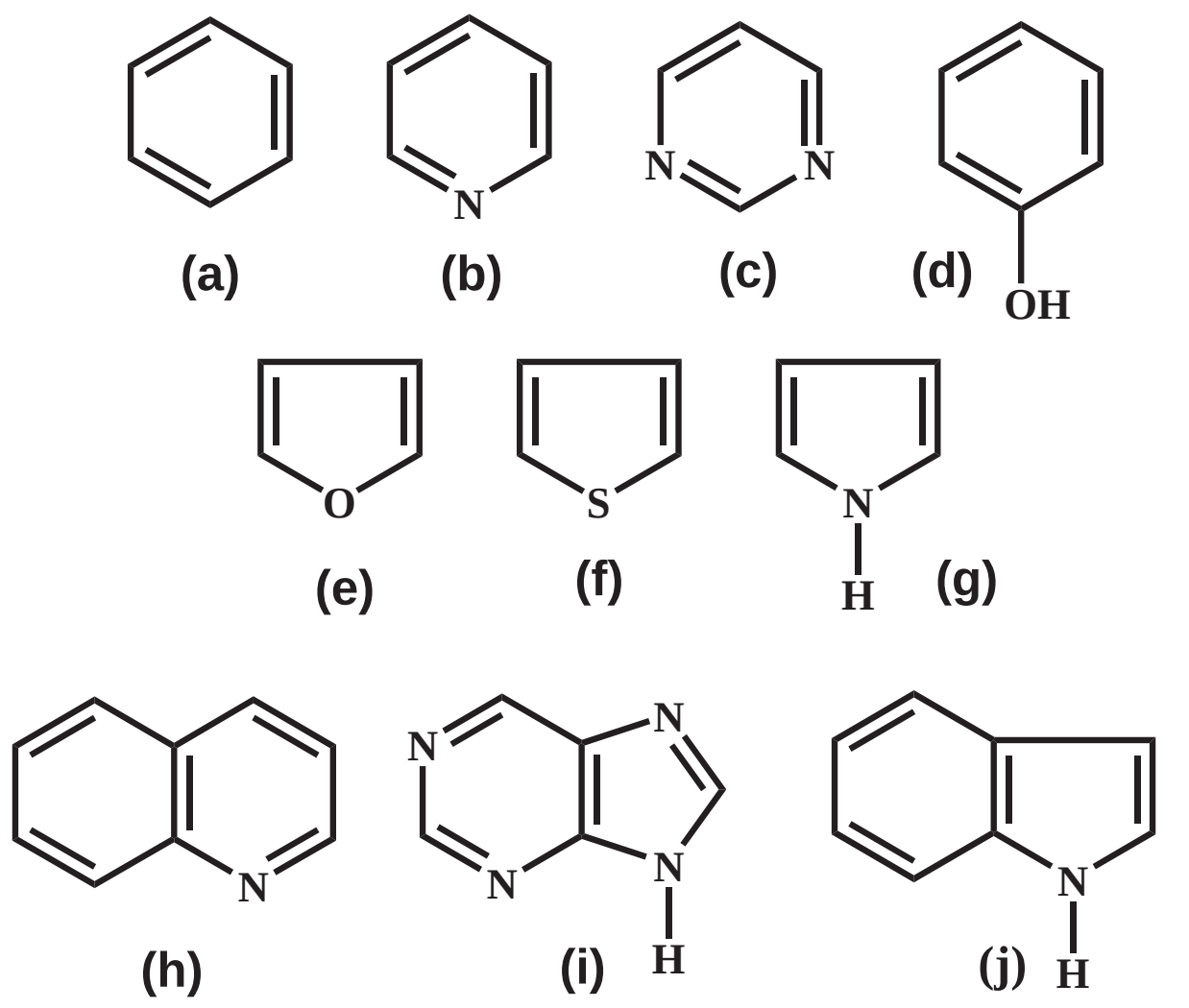

**Figure 7.** Several ring compounds containing nitrogen, oxygen and sulfur incorporated in heterocyclic system, which are used as building units in the design of molecular electronic materials. The molecules are named as (a) Benzene, (b) Pyridine, (c) Pyrimidine, (d) Phenol, (e) Furan, (f) Thiophene, (g) Pyrrole, (h) Quinoline, (i) Purine and (j) Indole.

**Table 6.** Functional substituents usually encountered in molecular electronic materials, especially in molecules for monolayer devices.

| Entry | Substituent (abbr.) | Formula |
|---|---|---|
| 1 | Acid (Ac) | $-COOH$ |
| 2 | Alcohol | $-OH$ |
| 3 | Amine | $-NH_2$ |
| 4 | Ethyne | $-C_2H_2$ |
| 5 | Ethene | $-C_2H_4$ |
| 6 | Ethyl (Et) | $-C_2H_5$ |
| 7 | Methoxy | $-OCH_3$ |
| 8 | Methylene | $=CH_2$ |
| 9 | Methyl (Me) | $-CH_3$ |
| 10 | Nitro | $-NO_2$ |
| 11 | Nitrile | $-CN$ |
| 12 | Thiol | $-SH$ |
| 13 | Thiolacetyl (AcS) | $-SCOOH$ |
| 14 | Trichlorosilane | $-Si(Cl)_3$ |
| 15 | Triethoxysilane | $-Si(OC_2H_5)_3$ |
| 16 | Trimethylsilane | $-Si(CH_3)_3$ |
| 17 | Vinyl | $-C_2H_3$ |

### *7.2.2.3 Approaches for Molecular Junction Fabrication*

Single molecule-scale memory devices and their integrated system are the long-term objects for the intense research in the field of molecular electronics. Experimental and theoretical efforts by numerous scientists and engineers confirmed that it is very promising to realized molecular-scale information memories. One of the biggest obstacles is how to fabricate addressable single molecular junctions with a higher reproducibility and durability.

To date, the researchers have developed many effective approaches to fabricate MOM junctions. Table 7 summarized the fabrication methods reported in literatures. From this table, we learn that: 1) the liquid metal droplet and crossed microwire [15] junction methods are simple, but the microscale junctions are neither stable nor addressable. 2) The techniques based on SPM [16], electrodeposited [17], and against nanoparticles [18,19] and mechanically breaking nanowires are powerful in studying molecular electronics, but they will be applicable only if the addressing problems can be solved. 3) The crossbar [6,20,21] and nanopore [22-24] approaches are applicable in the investigation of molecular devices with dimensions down to 0.1 micrometer, which is limited by the resolution of e-beam and photolithography. 4) It is suggested that SAM-based crossed nanowire junction be the most promising approach to meet the application requirements for nanoscale information memory from the aspects of junction size, addressing and reliability. It is worth more researching attentions.

**Table 7.** Summary and comparison of molecular junction fabrication approaches. [Adapted with permission from Ref. 5. Copyright (2009), World Scientific Publishing Co.]

| **Fabrication approach** | | | **Junction dimension** | **Addressable** | **Temperature. variation** | **Nanoscale** |
|---|---|---|---|---|---|---|
| Liquid metal droplet | Hg | | 50~200 $\mu$m | Not | Not | Not |
| | GaIn | | | | | |
| SPM tip | STM | STM tip | 1~5 nm | Not | Not | Yes |
| | | Nanoparticle coupled STM | 1~5 nm | Yes | Not | Yes |
| | CAFM | CAFM tip | 5~50 nm | Not | Not | Not |
| | | Nanoparticle coupled CAFM | 1~5 nm | Yes | Not | Yes |
| Crossbar | Magnetically controlled crosswire | | 0.5~5 $\mu$m | Not | Not | Not |
| | Lithography crossbar | | 2~50 $\mu$m | Yes | Yes | Not |
| | Stamp-printing crossbar | | | | | |
| | Crossed nanowire | | 1~5 nm | Yes | Yes | Yes |
| Etched hole | Etched hole plus nanotubes | | 2~10 $\mu$m | Yes | Yes | Not |
| | Nanopore | | 30~50 nm | Yes | Yes | Not |
| Against nanowire | Electromigrated | | 1~70 nm | Yes | Yes | Unknown |
| | Electrodeposited | | 1~50 nm | Yes | Yes | Unknown |
| | Mechanically broken | | 1~50 nm | Not | Not | Unknown |
| Against nanoparticle | Nanoparticle bridged | | 1~5 nm | Yes | Yes | Yes |

### *7.2.2.4 Device Characterizations*

The electronic information of molecular material can be deduced from direct device electrical performance like current-voltage characteristics. There are two regular methods to measure a molecular junction: 1) apply a current and measure voltage, or 2) apply a voltage then measure current. If the molecular junction has a higher conductance, the first technique generally yields a good result. On the other hand, if the device's resistance is very high, the second method is a better choice.

As shown in the schematic diagram of figure 8, the charge transport performance of a molecular junction depends not on the organic molecule itself but on the molecule/electrode interface(s). The influencing molecule-specific factors include molecular length, functional group/ligand, conformation and electronic structure etc. The work function values of metal electrodes are also an important element. For regular electrode materials, the work functions of GaIn, Al, Cr, Cu, ITO, Au and Ni are 4.15, 4.28, 4.5, 4.65, 4.95, 5.1 and 5.15, respectively.

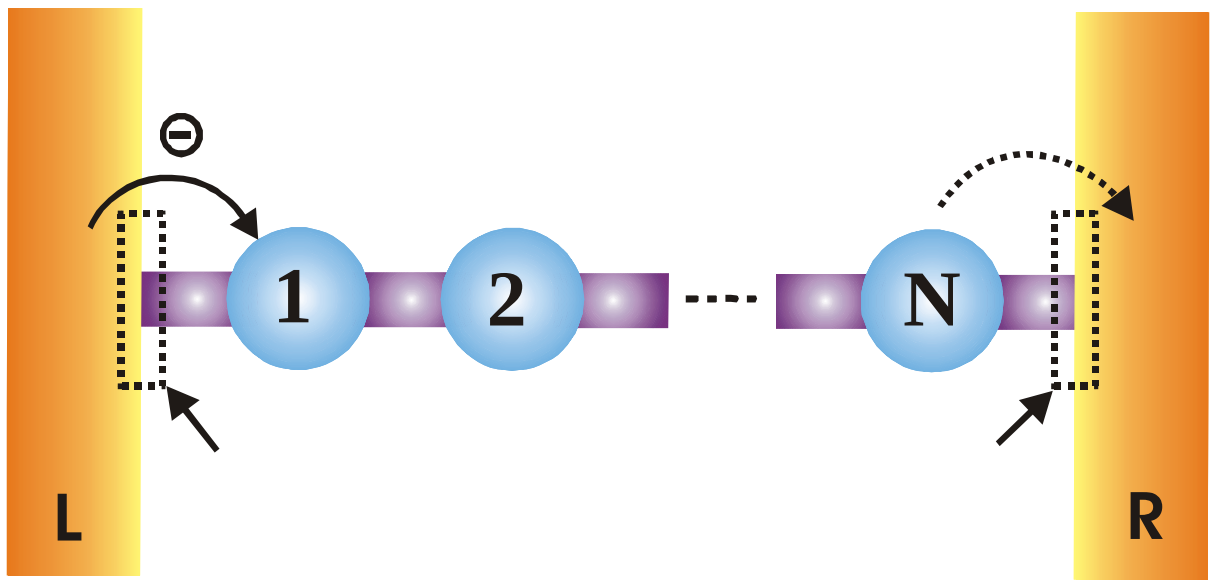

**Figure 8.** Schematic structure of a junction with one molecular wire contacted with left (L) and right (R) electrodes. The left and right electrode/molecule interfaces can drastically affect the junction's charge transport and performance. The molecular functional units are labeled from 1 to N.

Although there are some significant research results reported, it is still poorly understood about the dependence of junction switching performances on molecule-specific properties. To characterize the device performance of a molecular junction, the following factors need to be carefully examined:

1) The substrate effect, i.e. how the substrate property affect the junction performance.

2) The solvent-induced effect, if the junction is investigated under solution.

3) Active layer thickness and morphology effects.

4) Effects of oxygen and/or water in the measuring environments.

5) Electrode/molecule interfacial effects including contacting nature (ohmic or non-ohmic), charge transfer and electronic energy structure (Table 8 lists out the work function values for some metals used in molecular electronics).

6) Influence of optical, temperature, molecular self-heating and molecule-specific properties (such as length, ligand and conformation).

7) Finally, the device contamination, variability and reproducibility are also need to be under examination.

**Table 8.** Work function of metals used in molecular electronics. Note: the values are just for clean metal electrodes, it may drastically change upon contact with organic layers.

| **Electrode** | **GaIn** | **Al** | **Cr** | **Cu** | **ITO** | **Au** | **Ni** |
|---|---|---|---|---|---|---|---|
| Work function (eV) | 4.15 | 4.28 | 4.5 | 4.65 | 4.95 | 5.1 | 5.15 |

### 7.2.3 Switching Mechanisms

To effectively control and enhance the molecular switching behaviors at room temperature, it is required to understand the device dominant switching mechanism from the viewpoint of structure-property relationship. The investigation of the effect of chemical structure/functional groups on the device switching performance can yield a great deal of information about both the charge transport and switching mechanisms. The electrical and/or optical switching mechanism of a molecular junction may be resulted from many factors that at least include: 1) Metal Filament Formation; 2) Molecule-Electrode Interface Effect; 3) Charge Transfer Effect; 4) Charge Trapping-Detrapping (or Doping-Dedoping) Effect; 5) Redox Effect and 6) Molecular Conformation Change. Table 9 presents the comparison result for different molecular device systems operated under these mechanisms. It is indicated that the molecular systems with charge transfer and redox switching performance are promising for nanoscale information memory applications.

**Table 9.** Comparison of switching mechanisms of molecular memory systems.

| Switching mechanism | Molecule intrinsic | Reversible | Transition time | Stability |
|---|---|---|---|---|
| Metal filament formation | Not | Not | Long | Bad |
| Molecule/electrode interface | Not | Sometimes | Short | OK |
| Charge trapping-detrapping | Not | Yes | Short | OK |
| Conformation change | Yes | Yes | Short | Good |
| Charge transfer | Yes | Yes | Shorter | Good |
| Redox effect | yes | yes | Shorter | Better |

## 7.3 Redox Molecule-Based Memory

Recently, Blackstock et al reported the growth of pinhole-free thin films for a series of polyarylamine materials with redox gradient properties [25-27]. Except for their good hole transport characteristics [28-31], one unique aspect of the compounds in this family is their spatial relationship of the redox active moieties, using easy-oxidize group as the molecular "core" and harder-oxidize group as the "shell". The other special feature of this kind of material is their application for multilevel redox state information memory [32-34]. We investigated the fabrication of organic diodes and conduction properties of these materials in order to evaluate the correlation between the molecule-specific properties and the device performance [7,13,14]. We also introduce a simple method to directly measure the rectifying barrier between the molecular thin films and a removable GaIn liquid cathode. This simple approach can not only prevent the damage to the organic active layers, but also has the advantage of reproducibly measuring devices with active layer as thin as 50 nm thickness.

### 7.3.1 Charge Carrier Injection and Transport

Carrier injection into a traditional inorganic semiconductor is usually described by using either Fowler-Nordheim tunneling or Richardson-Schottky thermionic emission theories. The average mean free path of carriers in organic semiconductor is of the order of the molecular scale, which is far less than that of the inorganic one. Therefore, the charge carrier injection and transport in organic solids are more difficult than that in regular semiconductors. The carriers in organic solids have to overcome random energy barriers caused by disorder like polycrystal boundaries. However, we can roughly use the normal semiconductor theories to analyze the experimental results of organic thin film devices. As we discussed in the later section of this chapter, the following theoretical models and equations are usually used in analyzing the *I-V* curves of molecular junctions with active layer of either organic thin film or molecular monolayer.

**Richardson-Schottky thermionic emission** is based on lowering of the image charge potential by the external electrical field. In this model, the current density $J$ is given as a function of the external field $F$ in the form of

$$J = A^{*}T^{2}\exp\left(-\frac{\phi_B - \beta\sqrt{F}}{k_B T}\right) \qquad (1)$$

with the Richardson constant $A^{*} = 4\pi q m^{*} k_B^2 / h^3$, constant $\beta = \sqrt{q^3 / 4\pi\varepsilon\varepsilon_0}$ and the zero-field injection barrier $\phi_B$. While $q$ is elementary charge, $m^*$ effective electron mass, $k_B$ Boltzmann's constant, $h$ Planck's constant, $T$ temperature, $\varepsilon$ relative dielectric constant and $\varepsilon_0$ vacuum permittivity.

**Fowler-Nordheim tunneling** mechanism ignores Coulombic effects and considers mere tunneling through a triangular barrier into continuum states. In this model, the current density is given by

$$J = \frac{A^* q^2 F^2}{\phi_B \alpha^2 k_B^2} \exp\left( - \frac{2\alpha \phi_B^{3/2}}{3qF} \right) \qquad (2)$$

with constant $\alpha = \frac{4\pi\sqrt{2m^*}}{h}$. Unlike that of the Richardson-Schottky theory, there is no temperature dependence of the current density in this model. For the charge carrier transport in a thin film-based molecular junction, the conductance can be simply regarded as either space charge limited or injection limited. The readers can refer to some good review articles for further details.

### 7.3.2 Redox Materials, Thin Film and Property

Figure 9 shows the molecular structures of sixteen molecular compounds studied for redox dendrimeric family. Molecules **a** and **j**–**o** are designed in the lab of Prof. Blackstock, while the synthesis of **k**, **l**, and **n** has been previously reported [25-27]. Based on the molecular structure, the materials could be intentionally divided into three groups (1st group **a**–**c**, 2nd group **d**–**i**, and 3rd group **j**–**p**, respectively). Note that the moderate size and dendritic structures are the common characters for the molecules of the third group. As discussed later, this classification would be helpful in understanding the correspondence between the molecular structure and the device performance.

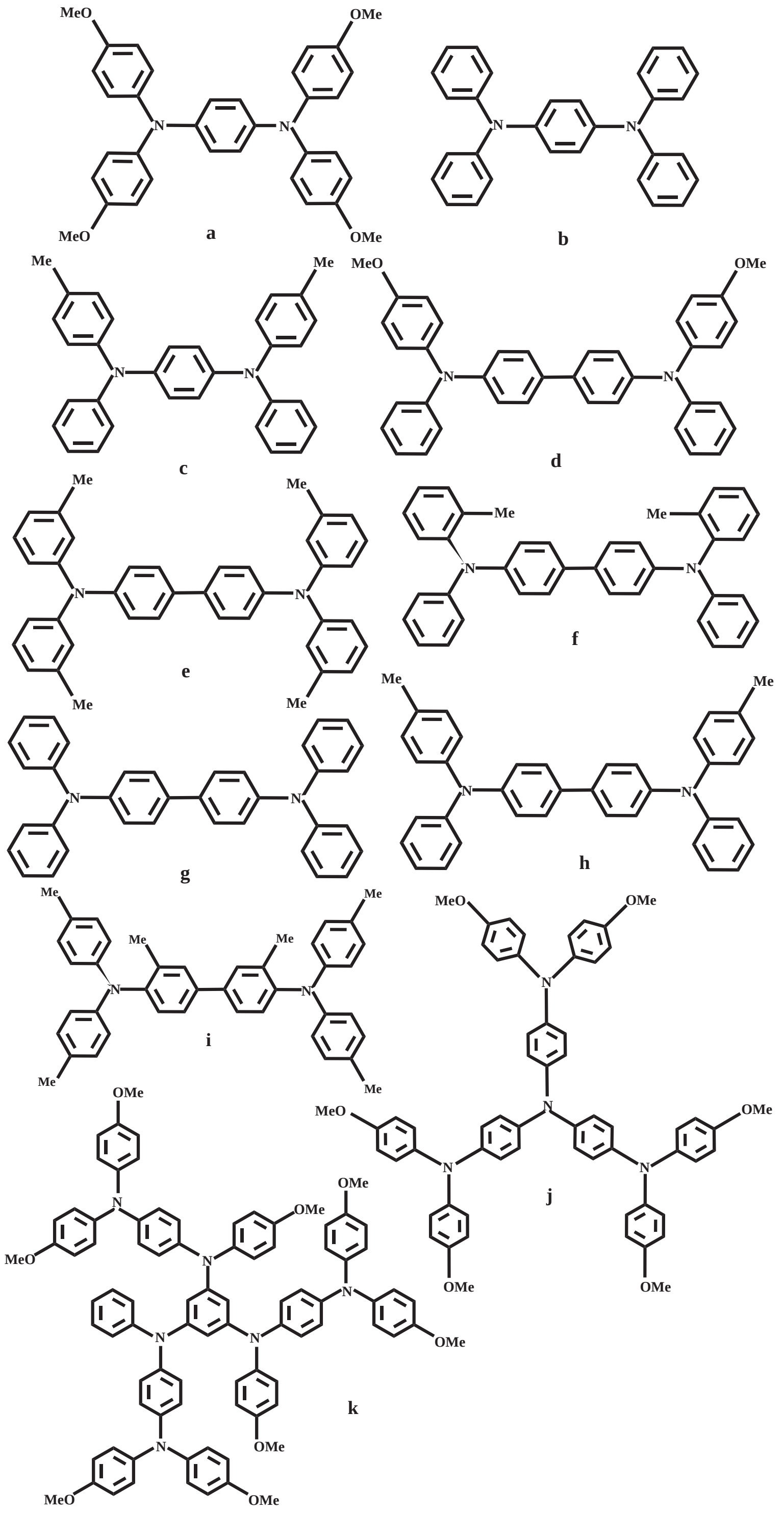

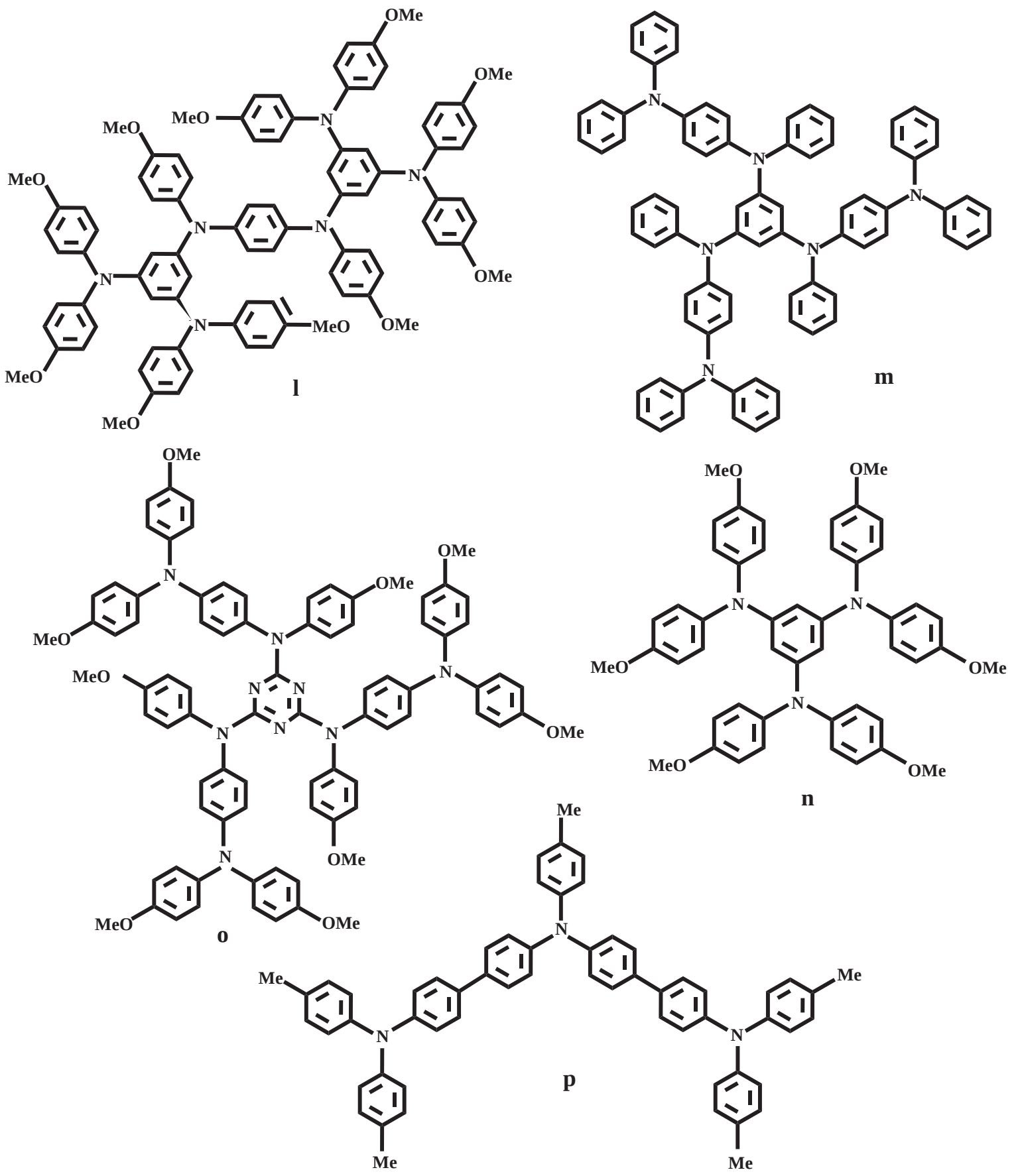

**Figure 9.** Chemical structure of a series of redox arylamine molecular materials. The first Group includes molecules **a, b** and **c**, the second group are molecules **d–i**, while the left redox arylamine molecules are classified as the third group. Here, molecule **i** is widely known as TPD, a hole transport materials used in organic light emitting diodes.

Rectification was observed in the organic diodes made from each molecular material. Figure 10 shows the representative I–V curves for six materials. Inset shows the schematic device structure and the measuring system used. Forward bias corresponds to holes emitted from the Ag anode and collected by the GaIn cathode. Negligible leakage current was measured below the turn-on voltage. Above the turn-on voltage, the current shows non-linear behavior. A preliminary analysis of these characteristics can be described by the space charge limiting conduction model.

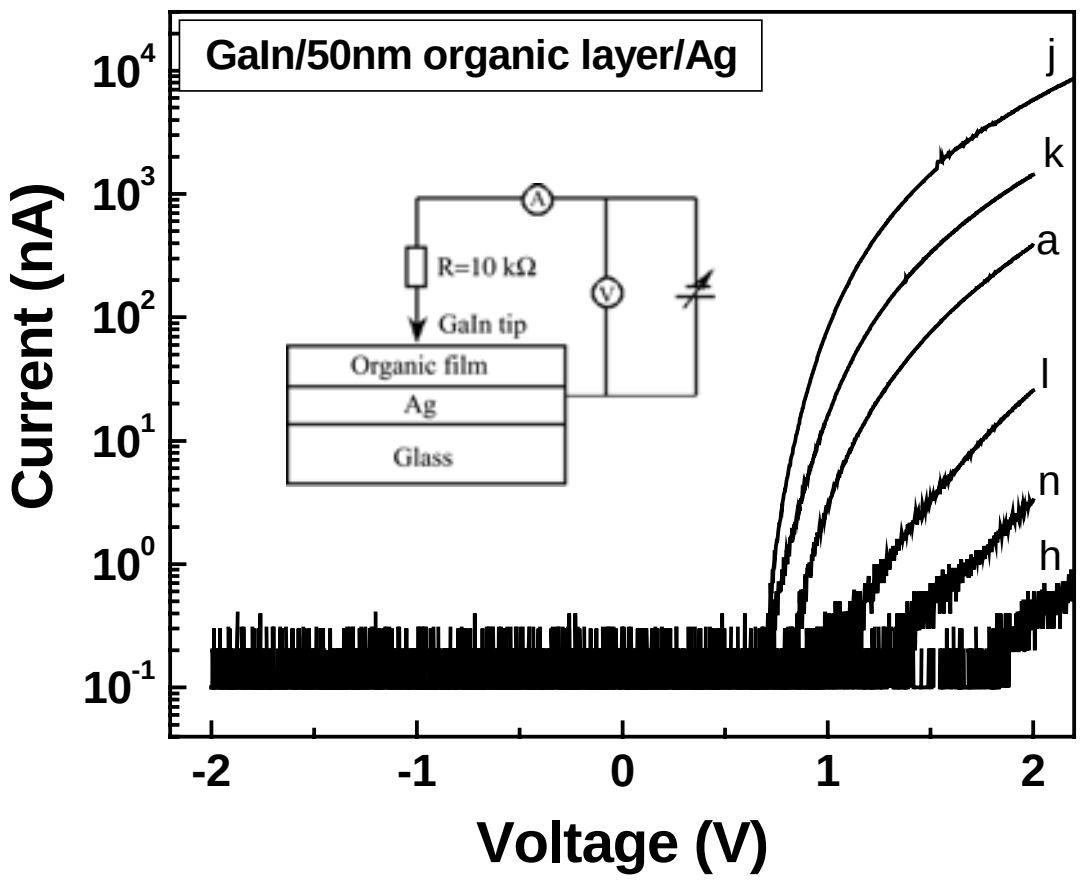

**Figure 10.** Representative I-V curves for six typical materials. It is clear that the turn-on voltage of different molecular thin films are drastically different although the active layer was kept in the same thickness. Inset shows the schematic device structure and measuring system used [13].

Typical tapping mode AFM images of several representative molecular thin films deposited on Ag surface are presented in Figure 11. We found that the film morphology and device performance are closely correlated with the molecular chemical structures. Briefly, the small molecular materials, with symmetry structures, tend to form rough film (molecules **a**–**c**) or smooth films either with lots of pinholes (molecules **e, f** and **i**) or with loose-packed film morphology (molecules **d, g** and **h**). Consequently, their devices show very poor performance as indicated by the low good device rates. In contrast, the films deposited from the materials with moderate molecular size and dendrimeric structures, usually show both smooth and close-packed characters. As a result, most of these devices exhibit very high good device rate. These results indicate that the morphology effect induced by the molecular structure plays an important role in the single carrier organic diodes [13]. The data is summarized in Table 10.

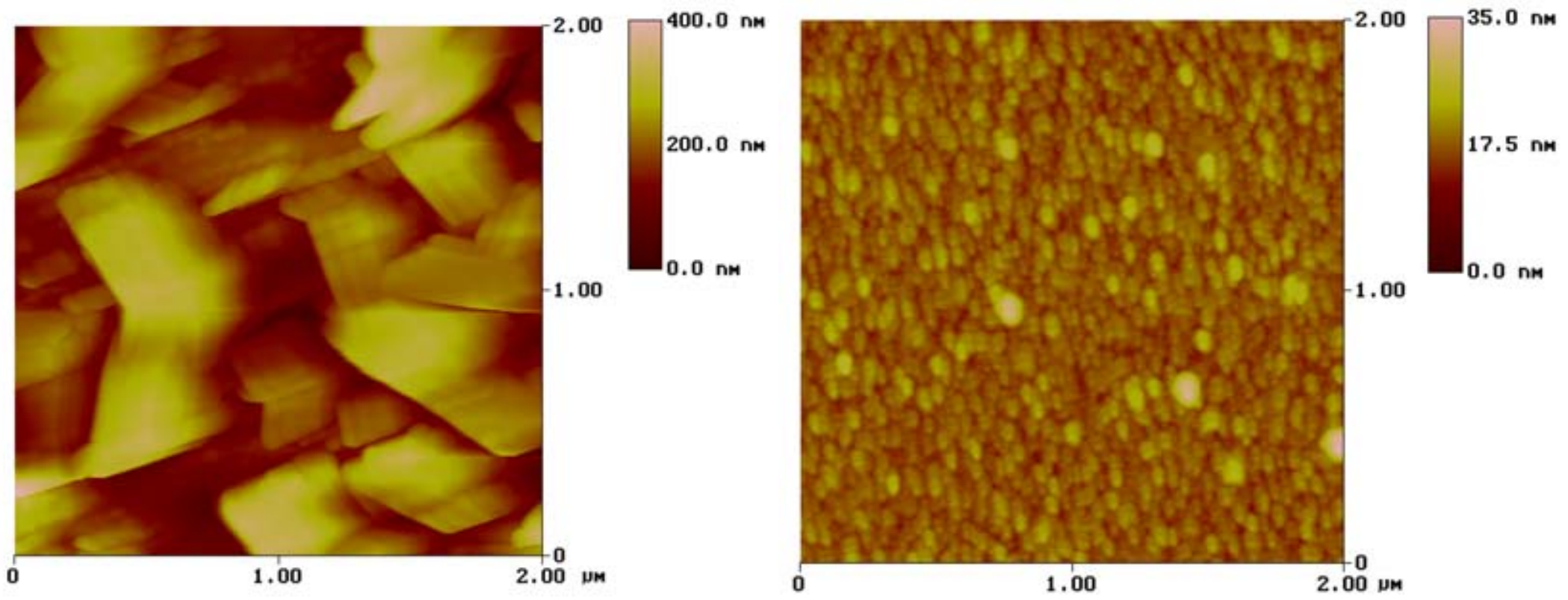

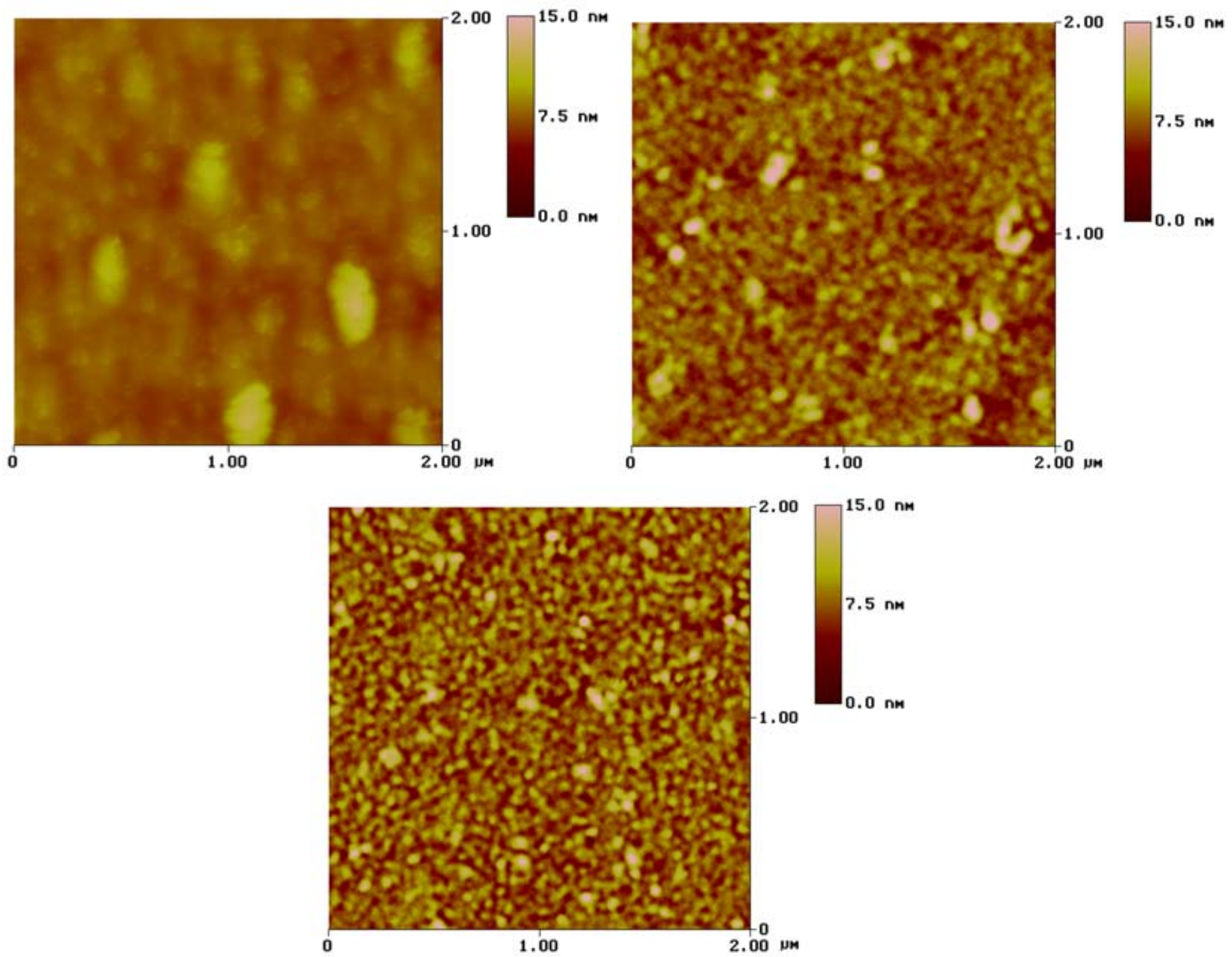

**Figure 11.** Tapping mode AFM images for the thin films of figure 9 molecules **a**, **h**, **n**, **k** and **l** (from top to down), respectively. The films are 50-nm-thick vapor deposited on Ag electrodes. Low molecular weight materials with symmetry structures tend to form rough and loosely packed films, while moderate molecular weight and dendrimeric materials usually show smooth and close-packed characters [13].

**Table 10.** Comparison of the molecular first oxidation potential, thin film morphology and electrical performance of GaIn/50 nm redox film/Ag junctions.

| Molecule | 1st oxidation potential (V vs. SCE) | Turn-on voltage (V) | Mean roughness (nm) | Good junction percent (%) |
|---|---|---|---|---|
| a | 0.46 | 0.79 ± 0.09 | 75 | Poor (< 10) |
| b | 0.56 | N/A | 80 | Very poor |
| c | 0.60 | N/A | 35 | Very poor |
| d | 0.76 | N/A | 1.2 | Very poor |
| e | 0.80 | N/A | 2.1 | Very poor |
| f | 0.81 | N/A | 3.9 | Very poor |
| g | 0.83 | N/A | 1.1 | Very poor |
| h | 0.85 | 1.80 ± 0.20 | 1.8 | Poor (< 25) |
| i | 0.9 | N/A | 1.0 | Very poor |
| j | 0.32 | 0.67 ± 0.09 | 1.7 | 80 |
| k | 0.45 | 0.71 ± 0.06 | 1.0 | 95 |
| l | 0.49 | 1.03 ± 0.11 | 1.4 | 85 |
| m | 0.64 | 1.52 ± 0.17 | 1.9 | 30 |
| n | 0.67 | 1.36 ± 0.09 | 1.1 | 85 |
| o | 0.71 | 1.53 ± 0.06 | 1.2 | 70 |
| p | 0.74 | 1.14 ± 0.06 | 1.8 | 90 |

It is interesting to compare our results with those of the multilayer devices reported in literatures. For example, molecule **p** exhibits high durability in multilayer devices, which has a good device rate in our diodes. Contrarily, the molecules (**b**, **c**, **g** and **i**), with poor good device rate in our case, show very low durability in the multilayer devices. Therefore, our results are well consonant with that of the multilayer devices. Moreover, it may suggest that single carrier organic diodes might provide a simple way to screen molecular materials for the study of practical multilayer devices. Several large compounds, with higher dendrimeric degree and bigger molecular size than that of molecule **k**, have been intentionally synthesized and measured. No reasonable results could be obtained due to material thermal decomposition during the thermal evaporation process.

The influence of anode work function on the device electrical performance was carefully investigated. As shown in figure 12, we characterized the electrical response of molecule **k**, exhibiting the best performance in these molecules, with using Ag, Au, Cu, and ITO anodes, respectively [references]. No apparent relationship between the anodes and the good device rate is observed. Independent of anode metals, the thickness dependence of the device turn-on voltage shows linear behavior too. The ordinate intercepts of the turn-on voltage at zero film thickness just agree with the work function difference between the GaIn cathode and the corresponding anodes. This result confirms that there are no evident interfacial effects in our single-carrier organic devices.

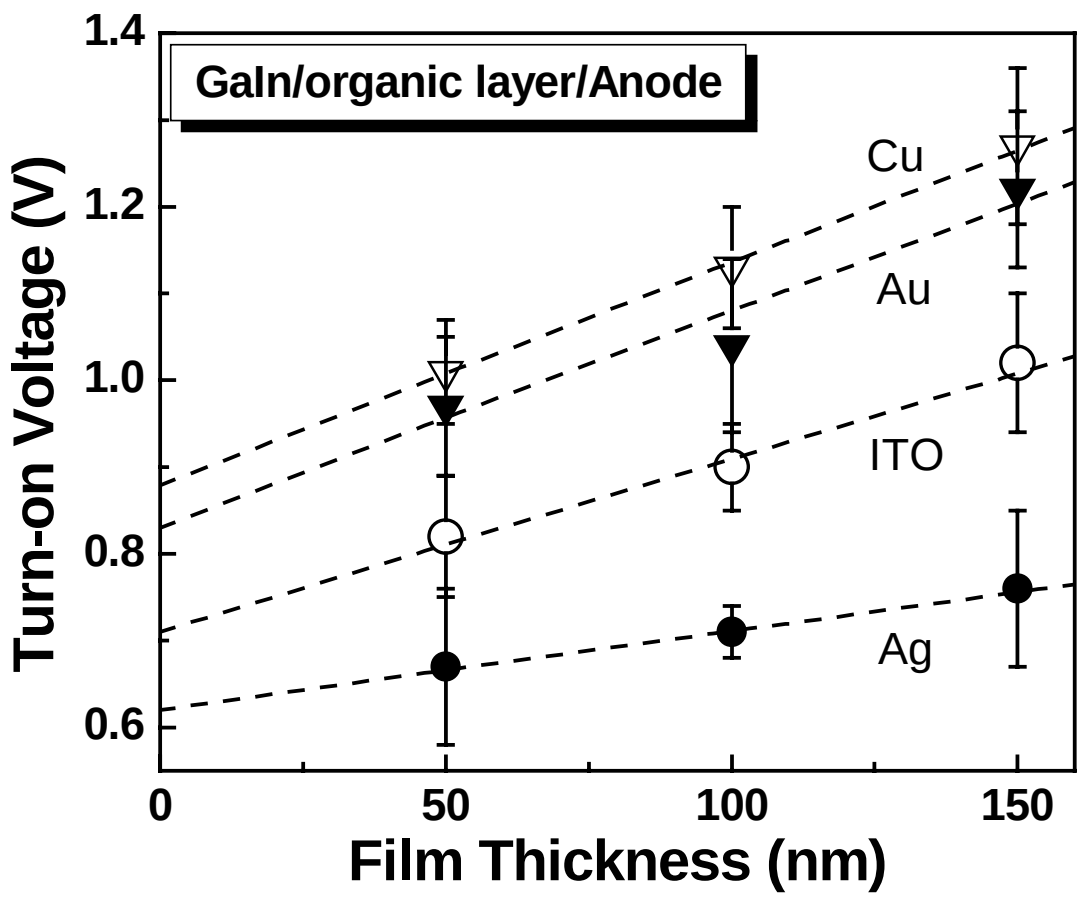

**Figure 12.** Effect of anode metal work function on the device turn-on voltage for molecule **k**.

The junction turn-on voltage ($V_t$) can be described as $V_t = \Delta W + kd$. Here, ΔW is the work function difference between the anode and cathode; $k$ is a constant related to the molecular properties; d is the thickness of the organic layer. In this formula, the first term represents the built-in potential, the second part is the voltage reduction consumed in the organic layer. When both electrodes and organic layer thickness are fixed, turn-on voltage could be tuned by changing the value of $k$. This result clearly demonstrates how the molecular properties could play a unique role in tuning the device turn-on voltage. On the contrary, when both the electrodes and molecular material are selected, turn-on voltage can only be lowered by reducing the layer thickness d. However, this will result in a high leakage current and low quantum efficiency.

### 7.3.3 Interfaces Between Redox Molecular Film and Electrode

The interfaces between metals and organic molecular thin films play a crucial role in the electrical and/or optical performance of molecular electronics, which include organic light emitting diodes, thin film transistors, solar cells, sensors and information memory devices. The metal/organic interfacial properties may be affected by various factors such as: 1) molecular structure and functional groups; 2) metal work function; 3) approaches of thin film and metal electrode deposition; 4) interface morphology; 5) charge transfer and buffer layers.

We previously studied the interface between arylamine thin films and various metal electrodes [14]. As shown in figure 13, it was found that there exist a charge transfer (CT) interface between metal electrode and arylamine molecular thin film when the molecule has a electron acceptor group like –CN. Moreover, the CT interface can be enhanced if the metal surface is pre-modified with a self-assembly monolayer of alkanethiol molecules, suggesting there exists a new CT interface between metal and SAM. The CT interface can be controlled through inserting a buffer layer between the metal and electron acceptor film. As a buffer layer materials, the molecule should have a very weak interaction with the metal electrode. The interfacial interactions between metal electrodes and organic materials with various functional groups usually follow the next, from weak to strong, sequence [35-45]:

$$-OCH_3 < -OH < -CH_3 < -Br < -CF_3 < -Cl < -CN < -F \qquad (3)$$

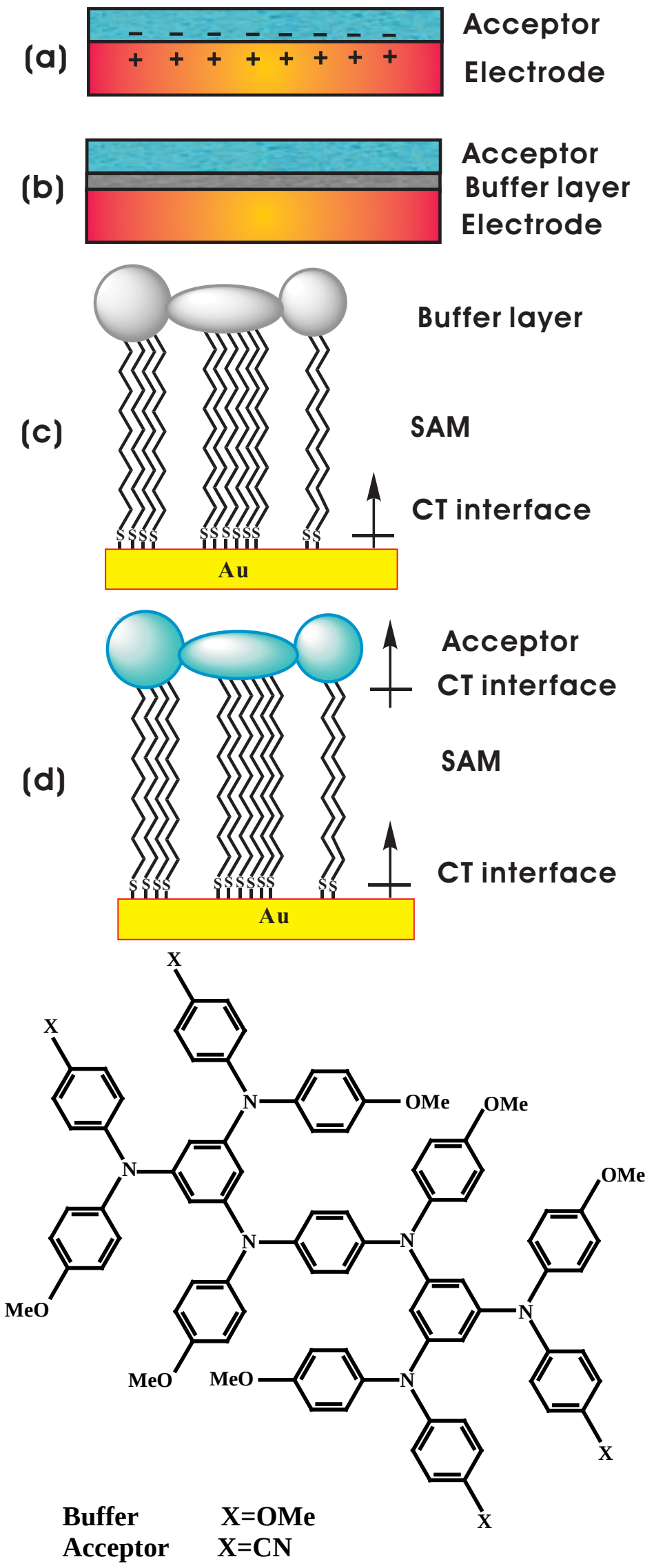

**Figure 13.** Schematic drawings show the charge transfer (CT) interfaces between redox molecular thin films with electron acceptor group (–CN) on substrates: (a) without buffer layer, (b) with buffer layer, (c) buffer layer on SAM and (d) acceptor film on SAM. Buffer layer means the molecular thin films made

from molecules without electron acceptor group. The molecular structure of the two kinds of materials is also shown.

Kahn et al reviewed the interface between metal electrode and organic thin films for a series of conjugated molecular materials with similar properties [46]. The chemical structure of the molecules is shown in figure 14, which are well studied materials used in organic light emitting diodes as charge transporting materials. The ionization energy (IE) for these materials is schematically shown in figure 15 to make a clear comparison. The IE was measured by ultraviolet photoemission spectroscopy (UPS), which is defined as the energy difference between the leading edge of the highest occupied molecular orbital (HOMO) and the vacuum level obtained from the photoemission cutoff. Figure 16 shows the comparison among metal work function, IE and electron affinity (EA) (i.e. HOMO and LUMO) positions of the molecular materials. LUMO is the molecular lowest un-occupied molecule orbital. The 'energy zero' is defined as the vacuum level. The IE and EA are determined by UPS and inverse photoemission spectroscopy (IPES). These data are very useful in studying the organic/metal interfaces such as charge transfer, dipole, work function changes, and energy barrier etc [38-45].

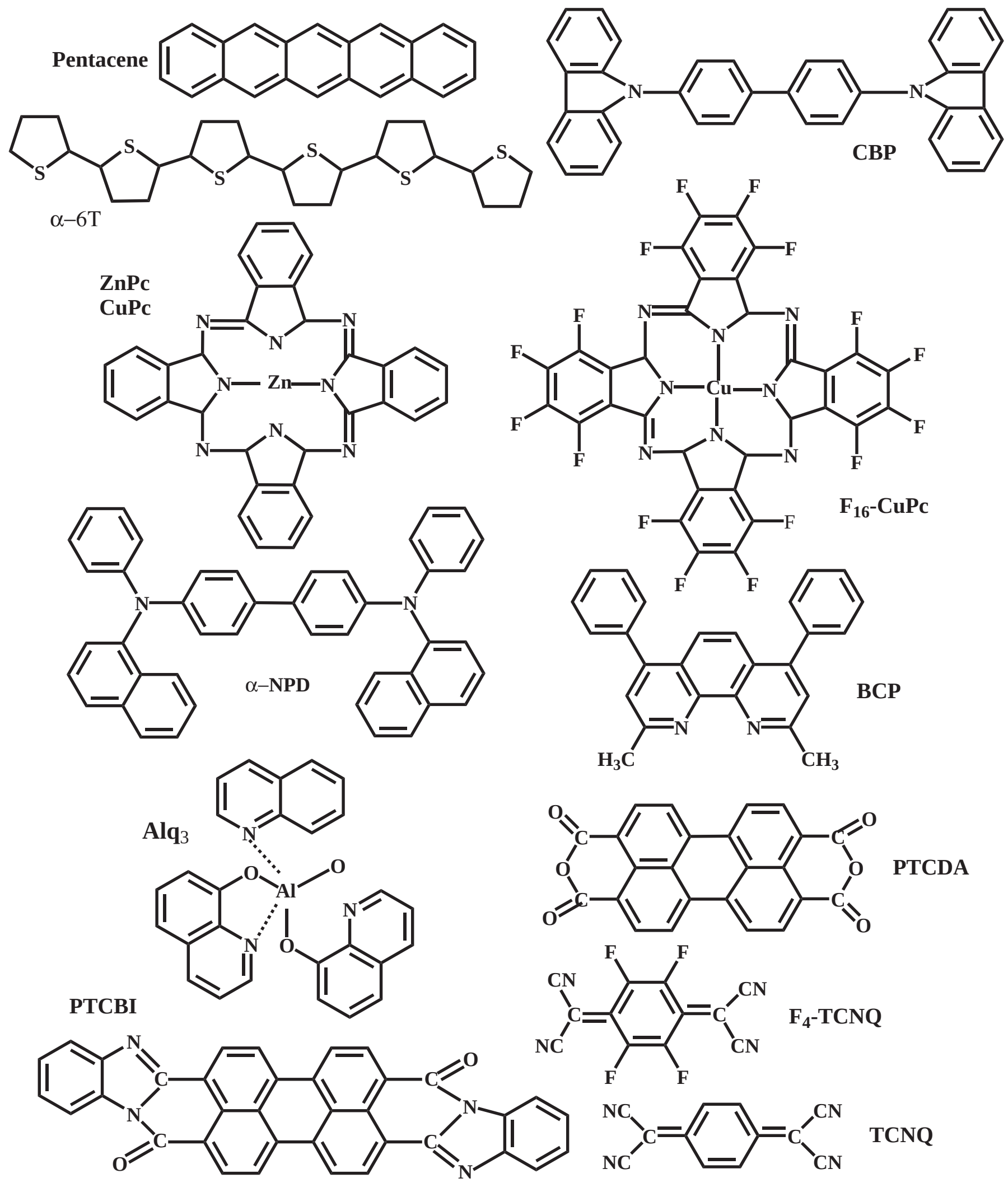

**Figure 14.** Chemical structure of the organic materials studied for understanding the electronic structure and electrical properties of interfaces between metals and pi-conjugated molecular thin films [46].

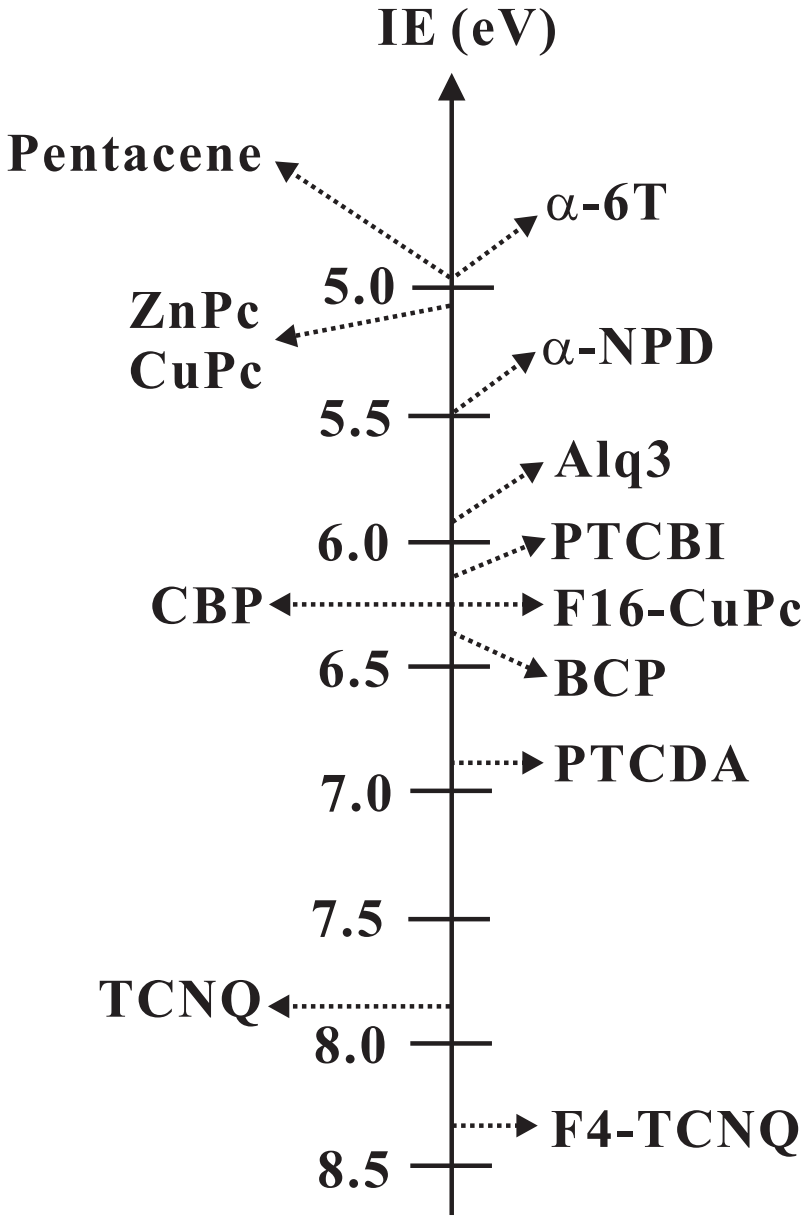

**Figure 15.** Comparison of the molecular ionization energy values of organic materials measured by ultraviolet photoemission spectroscopy (UPS). Here, IE is defined as the energy difference between the leading edge of the HOMO and the vacuum level obtained from the photoemission cutoff. The chemical structure of the molecules is given in a figure 14. [Adapted with permission from Ref. 46. Copyright (2003), Wiley Periodicals, Inc.]

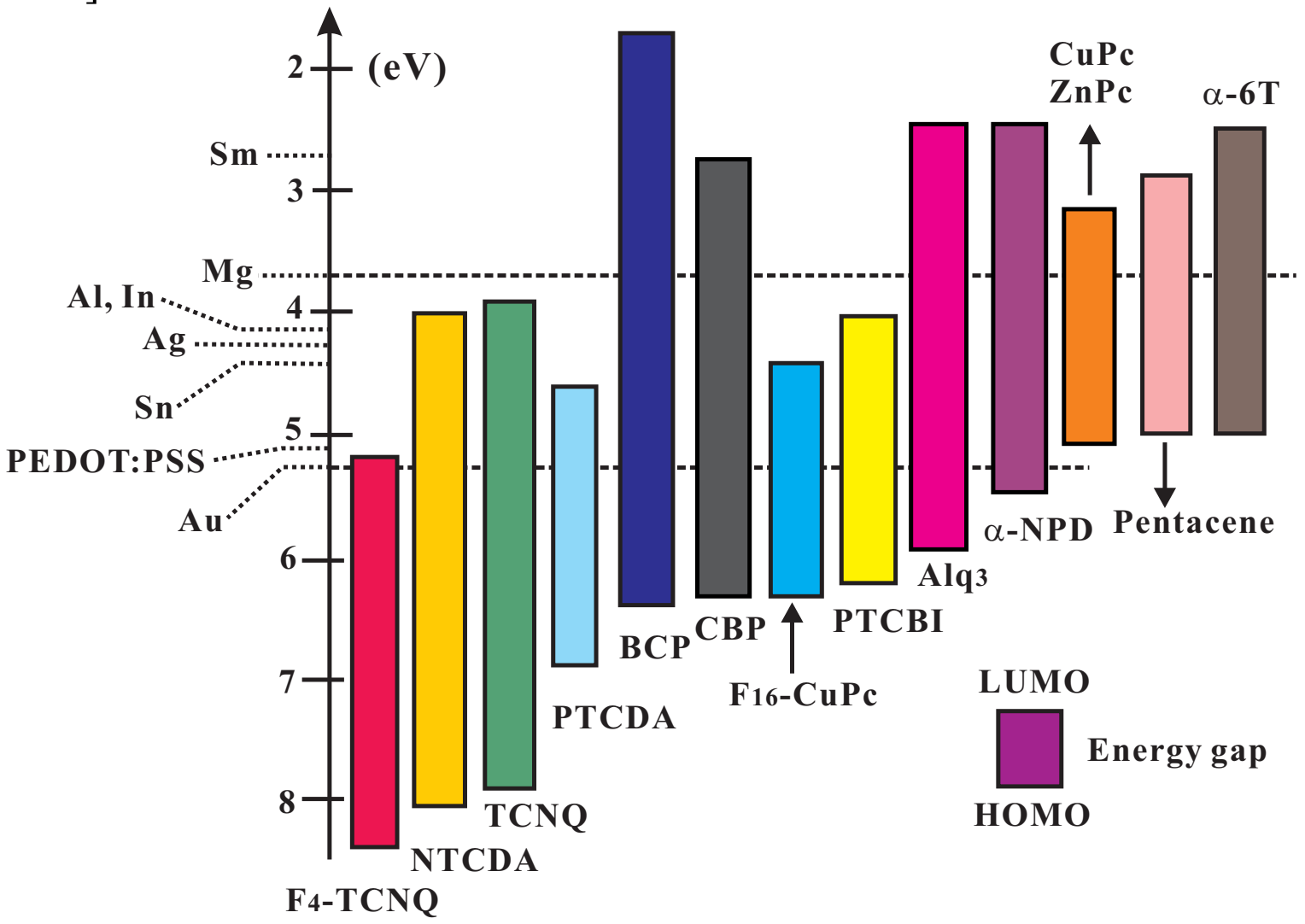

**Figure 16.** Comparison between metal work function, IE and electron affinity (EA) (i.e., HOMO and LUMO) positions of various molecular materials. The zero is defined as the vacuum level. The IE and EA are determined by UPS and inverse photoemission spectroscopy. [Adapted with permission from Ref. 46. Copyright (2003), Wiley Periodicals, Inc.]

### 7.3.4 Molecular Switching and Memory

Switchable redox molecular materials have drawn much researching attention owing to their special core-shell structure and multimode redox properties. Electrical switching and memory behavior was recently observed in the experiment of redox dendrimer sandwiched junctions with the structure of Ag/barrier layer/active layer/barrier layer/Ag [7]. Here, the barrier layers are molecular thin films that were vacuum vapor deposited from the molecule **i**, while the active layer was deposited from molecule **l** (see figure 9 for molecular structures). The molecular junctions were fabricated through a shadow mask with dimension of 1 $mm^2$. Control experiments confirmed that the switching mechanism is due to charge trapping in the redox molecules sandwiched by more insulated barriers. The switching threshold voltage (0.3 to 0.6 V) and on/off ratio (as high as 1000) depend on both the thickness and the ratio of the active and barrier layers. This work demonstrates the possibility of using solid-state redox thin films as information memory media.

As shown in figure 17(b), Wakayama et al. [47] observed optical switching in molecular multiplayer device with the structure of Au/ porphyrin-based molecues/Si (100), where the organic molecules are inserted into silicon oxide matrix layer. Their experiment was conducted under ultra-high vacuum conditions. Molecules of porphyrin derivative, tetrakis-3,5,di-butylphenyl-porphyrin, were deposited from an effusion cell and were sandwiched between two silicon dioxide layers (with total thickness of 5 nm). The current-voltage characteristics were measured at low temperature of 5 K. The light wavelength is 430 nm with intensity of 13 $\mu W/mm^2$ irradiated from a Xe lamp. The top gold electrode was deposited through shadow mask with thickness of 12 nm and transparency of 60%. The results indicated that the device current-voltage curves behaved obviously a Coulomb staircase originating from single-electron tunneling. The Coulomb staircase can be reversibly switched on/off by a threshold voltage of 300 mV under optical illumination light/dark conditions. It was suggested that the possible switching mechanism is photoinduced charge carriers trapping-detrapping at energy levels at the $Si/SiO_2$ interface. Such interfacial change will influence the junction tunneling parameters and thus result in optical switching. Liu et al. also reported their studies on the charge trapping/detrapping of photoconductive Zinc porphyrin film [48].

In a recent work [49], Chotsuwan and Blackstock reported on the observation of charge switching of small domains of polyarylamine isolated in polymethylmethacrylate (PMMA) through co-spin coating on SiO2/Si substrate (see figure 17(c)). The experiment was conducted under ambient conditions by using ambient conducting atomic force microscopy/Kelvin probe microscopy writing-reading technique. The thickness of the molecule mixed with PMMA is about three nm, while the thermal oxide layer of silicon is 25 nm. The device structure is Au-tip/molecules in PMMA/$SiO_2$-Si that is very similar to that of figure 17(b). It was shown that the thin layer could be charged with a threshold voltage of 6 V. This value is much large than that of the molecular first oxidation potential measured from cyclic voltammetry. The higher the amine content of the mixed film is, the easier of the film can be charged. Increasing the thickness of silicon dioxide layer can decrease the surface potential decay rate, suggesting reduction of the molecular charge by electrons from the silicon substrate. Moreover, they observed stepped discharge behavior, indicating that multi-mode redox switching happened in the experimental process. However, this method has the disadvantages of: 1) unknown distribution of the redox molecules in the PMMA and 2) difficulty to realize sub-10 nm scale information storage.

A molecular approach was described for information storage application that involves the usage of porphyrin derivative monolayer self-assembled on Au microelectrode as the memory storage element [50]. As shown in figure 17(d), the experiments were conducted in dried, distilled $CH_2Cl_2$ containing 0.1M $Bu_4NPF_6$ on a two-electrode potentialstat with a 5 MHz bandwidth. A set of four zinc prophyrins were examined, with each molecule bearing three mesityl groups and one S-acetylthio-derivatized linker with structure of 1-[AcS-$(CH_2)_n$]-4-phenylene (n=0,1,2 or 3). It was shown that multimode information storage could be realized through multiple oxidation states (neutral, monocation and dication) of the porphyrin molecules. The charge retention time is in the scale of hundreds of seconds and the redox process can be cycled thousands of times under ambient conditions. However, this approach involves the using of electrolyte and organic solvent, which is unfavorable for the practical applications of molecular scale information storage.

To avoid the problems mentioned above, we suggest a promising approach, STM tip-based information storage, on solid-state redox molecular monolayer with multilevel memory states [51, 52]. This method can easily reach a resolution of sub-5 nm scale. Experiments on similar molecular materials had been extensively used [53-55]. Moreover, we can easily add some special groups or ligand on the target molecules such as azobenzene [56-58], spiropyran/spirooxazine [59], and cyclohexadiene [60]. These functional groups will enable the molecules with optoelectronic switching properties that are very crucial for application in devices such as photodetector, photovoltatics and optically gated single molecule transistors.

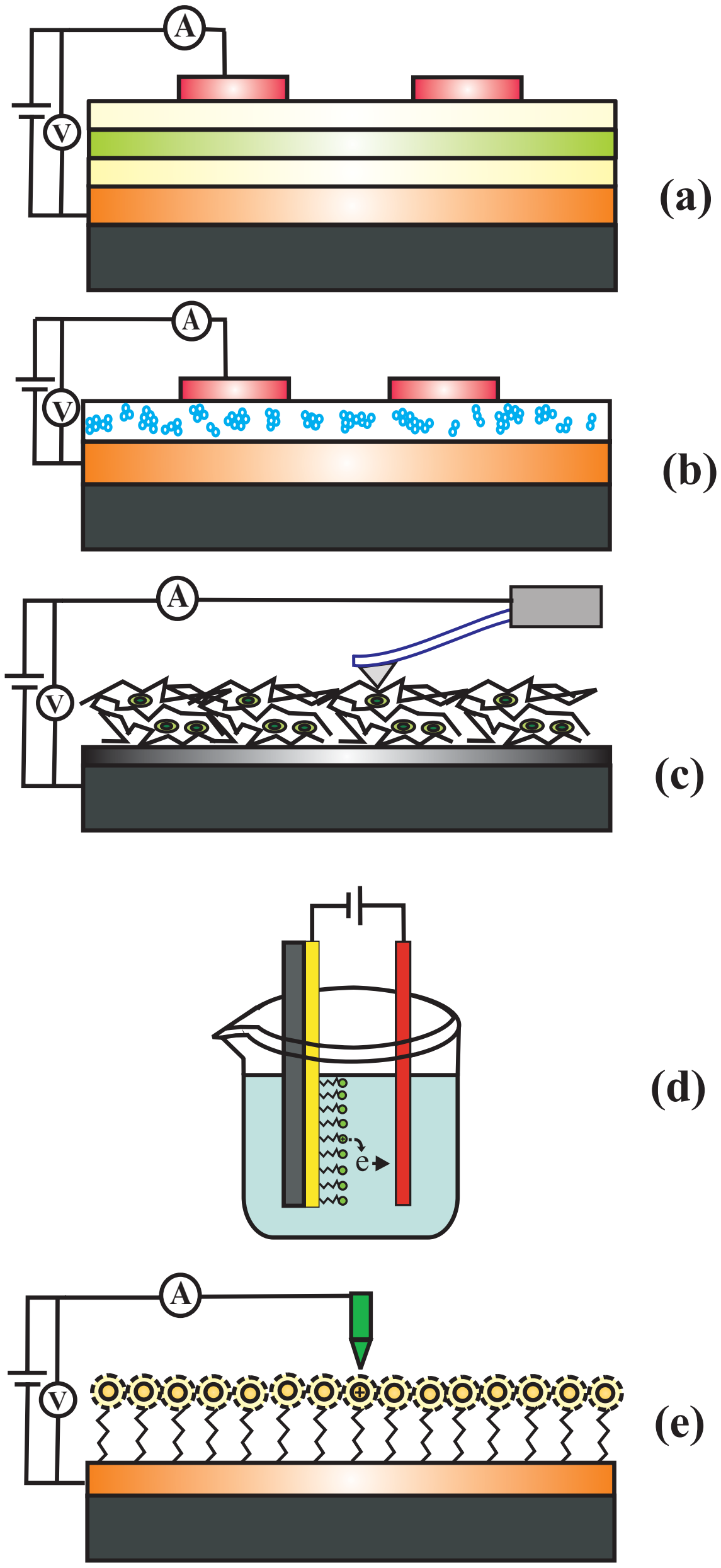

**Figure 17.** Schematic experimental setups for charge storage or electrical switching on redox dendrimeric thin films or monolayers. (a) Molecular junction of Ag/TPD/4AAPD/TPD/Ag [7], (b) Au/Molecule in $SiO_2$/Au [47], (c) Au-tip/Molecules in PMMA/$SiO_2$/Si [49], (d) Au/SAM/electrolyte [50] and (e) STM tip/Molecule/Au [53-55].

## 7.4 Self-Assembly Monolayer-Based Memory

In general, to realize molecular-scale data storage, we need to thoroughly understand the structure-property relationships of molecule, active layer and device. The goal of this section is to address the next questions: 1) what kinds of SAMs molecular materials are suitable for nanoscale memory applications? 2) How such monolayers can be patterned? 3) What is the structure-property relationship? 4) How can their information storage characteristics be greatly enhanced to function at room temperature?

### 7.4.1 Self-Assembly Monolayer and Property

For self-assembled monolayer based single molecular junctions, the charge transport properties and thus device performance may be drastically modified by factors of molecular structure or ligand [61], light illumination, thermal effect, and stress [62]. In fact, the electron tunneling in a single-molecule junction can be drastically varied even after a very small change of the molecule from the aspects of charge transfer, vibration, rotation, defects and/or conformation [63]. These effects must be taken into account in the investigation of molecular electronics, especially in the design and study of nanoscale memory devices [64].

#### *7.4.1.1 Charge Transport in Molecular Monolayer Junctions*

To discuss the charge transport mechanisms of monolayer junctions, Simmons model is simplest equation used to analyze the electron tunneling behavior through a rectangular barrier in the metal-insulator-metal junctions [65]. The equation of Simmon's model is

$$I = \frac{Ce}{4\pi^2\hbar d^2}\left\{\left(\phi-\frac{eV}{2}\right)\exp\left[-\frac{2(2m)^{1/2}}{\hbar}\alpha\left(\phi-\frac{eV}{2}\right)^{1/2}d\right]-\left(\phi+\frac{eV}{2}\right)\exp\left[-\frac{2(2m)^{1/2}}{\hbar}\alpha\left(\phi+\frac{eV}{2}\right)^{1/2}d\right]\right\} \quad (4)$$

where $e$ is the charge of an electron, $\hbar$ is Plank's constant divided by $2\pi$, $d$ is the tunneling distance (or wire length in the case of molecular junctions), $\phi$ is the barrier height, $V$ is the bias voltage applied between the electrodes, and $m$ is the mass of an electron. $C$ is the proportionality constant. $\alpha$ is a unitless adjustable parameter used in fitting. At lower bias voltage [66], the above equation can be approximated to $I = I_0 \exp(-\beta d)$ with conductance decay constant $\beta$ described as

$$\beta = \frac{2(2m)^{1/2}}{\hbar}\alpha(\phi)^{1/2} \quad (5)$$

This simplification has been widely used in characterizing molecular monolayers with techniques of CAFM and STM tip-based molecular junctions. The decay constant $\beta$ can serve as an indicator about the electrical properties of the molecules in the MOM junction. Keep in mind that, unlike that of the molecular junction in CAFM approach, there exists a small air gap between the STM tip and the molecular monolayer in the STM tip-based method [67]. Although the current-voltage characteristic of Simmon's model is independent of temperature, the situation may be extremely complicated in practical junctions, especially in high bias voltage regions [68].

Datta et al calculated the current through molecular junction of STM tip/α,α′-xylyl dithiol molecule/Au substrate by using the standard scattering theory of charge transport [69]. After comparing with the experimental results, Datta et al pointed out that we need to develop a more complete theory that includes other details such as the correct self-consistent potential profile or the structure in the density of states in the contacts. Later, Gonzalez et al modeled the influence of bridge electronic defects (site substitutions) and weak links (local weak bonds) on the potential profile of the molecular junction [70]. The potential is determined self-consistently by solving Poisson and Schrodinger equations simultaneously. Their calculated results show some asymmetry and rectification effects as that observed in real chemically modified molecular junctions. Lehmann et al theoretically investigated the incoherent charge transport through molecular wire junctions with the presence of Coulomb interactions [71]. The current for spinless electrons is determined in the limit of strong Coulomb repulsion. It is indicated that the voltage profile along the molecular wire crucially influences the dependence of the current on the wire length. Blocking effect is found upon the inclusion of the spin degree, which depends both on the interaction strength and on the number of the wires contributing to the current.

These works further remind us again that the charge transport of molecular junctions is very complicated due to lots of factors such as metal/molecule contacts, substitutions, chemical conformations and external effects. These will be further discussed in the next experimental sections.

### *7.4.1.2 Thiolated Self-Assembly Monolayers*

Salomon et al reviewed the electron transport measurements on organic molecules self-assembled gold surface [72]. Table 11 summarized the comparison results of sigma-bonded alkanethiol molecules investigated by using different junction approaches. Obviously, the current density values are drastically different for the same molecular structure with various investigation methods and/or under different measuring conditions. The SPM tip-based results are usually higher orders of magnitude than that of the others. It follows the trend of the less molecules within junction, the higher of the current through per-molecule at the given bias voltage. This can be ascribed to the exclusion of the intermolecular interactions in these measurements because of the nanoscale size and limited number of molecules within the junctions. This table highlights the importance of the chemically well-defined contacts and controlled numbers of molecule of the molecular junction.

The charge transport through pi-bonded conjugated molecular wires were also compared and summarized in Table 12 [72]. The chemical structure of the molecules is shown in figure 18. The absolute current per molecule of given length at a given bias, higher currents were obtained experimentally for conjugated pi-bonded molecules than for saturated ones with comparable length. Such observations agreed with that of the theoretical calculated results. From this table, it is also suggested that the electrode-molecule interface, the molecular conformation, and the molecular energy gap between HOMO and LUMO all play very critical roles for the charge transport of the molecular junctions.

**Table 11.** Comparison of current per molecule (*I*) through saturated sigma-bonded single molecules and monomolecular layers sandwiched between two electrodes measured by a variety of different experimental approaches at two different voltages (0.5 and 0.2 V). In the table, ‘Gap’ represents the molecular HOMO-LUMO energy gap and ‘Length’ is calculated molecular length. [Adapted with permission from Ref. 72. Copyright (2003), Wiley-VCH Verlag GmbH & Co. KGaA. Weinheim.]

| Entry | Junction | *I* at 0.5 V (pA) | *I* at 0.2 V (pA) | Gap (eV) | Area (No. of Mol. ) | Length (Å) | Method (force) |
|---|---|---|---|---|---|---|---|
| 1 | Au/Vacuum/Au | N/A | N/A | N/A | 0.2 $nm^2$ | 16 | Simmon’s model |
| 2 | Au-S-C8/Au | 30 | 12 | ~7 | 25 $nm^2$ (~100) | 11 | CAFM (2 nN) |
| 3 | Au-S-C8/Au | 0.035 | 0.013 | ~7 | 10 $nm^2$ (~40) | 11 | CAFM (6 nN in solvent) |
| 4 | Au-S-C8-S-Au | 1400 | 310 | ~7 | Single | 12 | CAFM (6nN in solvent) |
| 5 | Au-S-C8-S-Au | 1300 | 520 | ~7 | 25 $nm^2$ (~100) | 12 | CAFM (2 nN) |
| 6 | Au-S-C8-S-Au | 15000 | 2500 | ~7 | Single | 12 | Pico-STM |
| 7 | Au-S-C10/Au | 0.007 | 0.0002 | ~7 | 10 $nm^2$ (~40) | 14 | CAFM (6nN in solvent) |
| 8 | Au-S-C10/Au | 5.0 | 2.0 | ~7 | 25 $nm^2$ (~100) | 14 | CAFM (2 nN) |
| 9 | Hg-S-C10/p-Si | N/A | 6 | ~7 | 0.002 $cm^2$ (~ 1012) | 14 | Hg droplet |
| 10 | n-Si-C10/Hg | N/A | 2 | ~7 | N/A | 13 | Hg droplet |
| 11 | Au-S-C10-S-Au | 800 | 200 | ~7 | Single | 14 | CAFM (6 nN in solvent) |
| 12 | Au-S-C10-S-Au | 1200 | 370 | ~7 | Single | 14 | Pico-STM |
| 13 | Au-S-C12/Au | 0.2 | 0.1 | 7 | 1600 $nm^2$ (~6400) | 16 | Thermal deposition (nanopore) |
| 14 | Au-S-C12/Au | 0.5 | 0.2 | 7 | 25 $nm^2$ (~100) | 16 | CAFM (2 nN) |
| 15 | Hg-S-C12/p-Si | N/A | 0.6 | ~7 | 0.002 $cm^2$ (~ 1012) | 16 | Hg droplet |
| 16 | n-Si-C12/Hg | N/A | 0.5 | ~7 | N/A | 15 | Hg droplet |
| 17 | Au-S-C12/(Pt/Ir) | 0.55 | - N/A | ~7 | 0.25 $nm^2$ | 16 | STS |
| 18 | Au-S-C12-S-Au | 40 | 10 | ~7 | 250 $nm^2$ (~1000) | 17 | Crossed-wires |

**Table 12.** Comparison of currents through conjugated pi-bonded single molecules and monolayers sandwiched between two electrodes measured by a variety of different experimental approaches at two different voltages. In this table, 'Gap' represents the molecular HOMO-LUMO energy gap, Length is calculated molecular length. The molecular structure is shown in figure 19. [Adapted with permission from Ref. 72. Copyright (2003), Wiley-VCH Verlag GmbH & Co. KGaA. Weinheim.]

| Entry | Junction | Molecule | I at 0.5 V (pA) | I at 0.2 V (pA) | Gap (eV) | Area (No. of Mol.) | Length (Å) | Method (force) |
|---|---|---|---|---|---|---|---|---|
| 1 | Au-S-phenyl-S-Au | a | 1300 | 300 | ~5 | Single | 6 | Mechanical break |
| 2 | Au-S-phenyl-S/(Pt/Ir) | a | 110 | 30 | ~5 | 0.25 $nm^2$ single | 6 | STS |
| 3 | Au-S-C-phenyl-C-S/(Pt/Ir) | b | 560 | 130 | ~5 | 0.25 $nm^2$ single | 8 | STS |
| 4 | Au-S-biphenyl-S-Au | c | 36 | 4 | ~5 | Single | 11 | CAFM (12 nN) |
| 5 | Au-S-biphenyl/Ti | d | 300 | 7 | ~5 | 700 $nm^2$ (2800) | 10 | Thermal deposition (nanopore) |
| 6 | Au-S-molecules/Au | e | N/A | 500 | ~5 | 25 $nm^2$ (~100) | 11 | CAFM (2 nN) |
| 7 | Au-S-terphenyl/Au | f | N/A | 130 | ~5 | 25 $nm^2$ (~100) | 15 | CAFM (2 nN) |
| 8 | Au-S-OPV-S-Au | g | 1000 | 500 | 3.1 | 250 $nm^2$ (~1000) | 19 | Crossed wires |
| 9 | Au-S-OPE-S-Au | h | 500 | 200 | 3.5 | 250 $nm^2$ (~1000) | 20 | Crossed wires |
| 10 | Au-S-OPE-S-Au | h | 10 | 4 | 3.5 | Single | 20 | CAFM (6 nN in solvent) |
| 11 | Au-S-molecule-S-Au | i | 14× 104 | 4× 104 | 3.5 | Single | 20 | Mechanical break |
| 12 | Au-S-molecule-S-Au | i | 3× 104 | 2500 | 3.3 | Single | 20 | Electromigrated |
| 13 | Au-S-caroteno dithiol-S-Au | j | 100 | 40 | 2.4 | Single | 32 | CAFM (6 nN in solvent) |

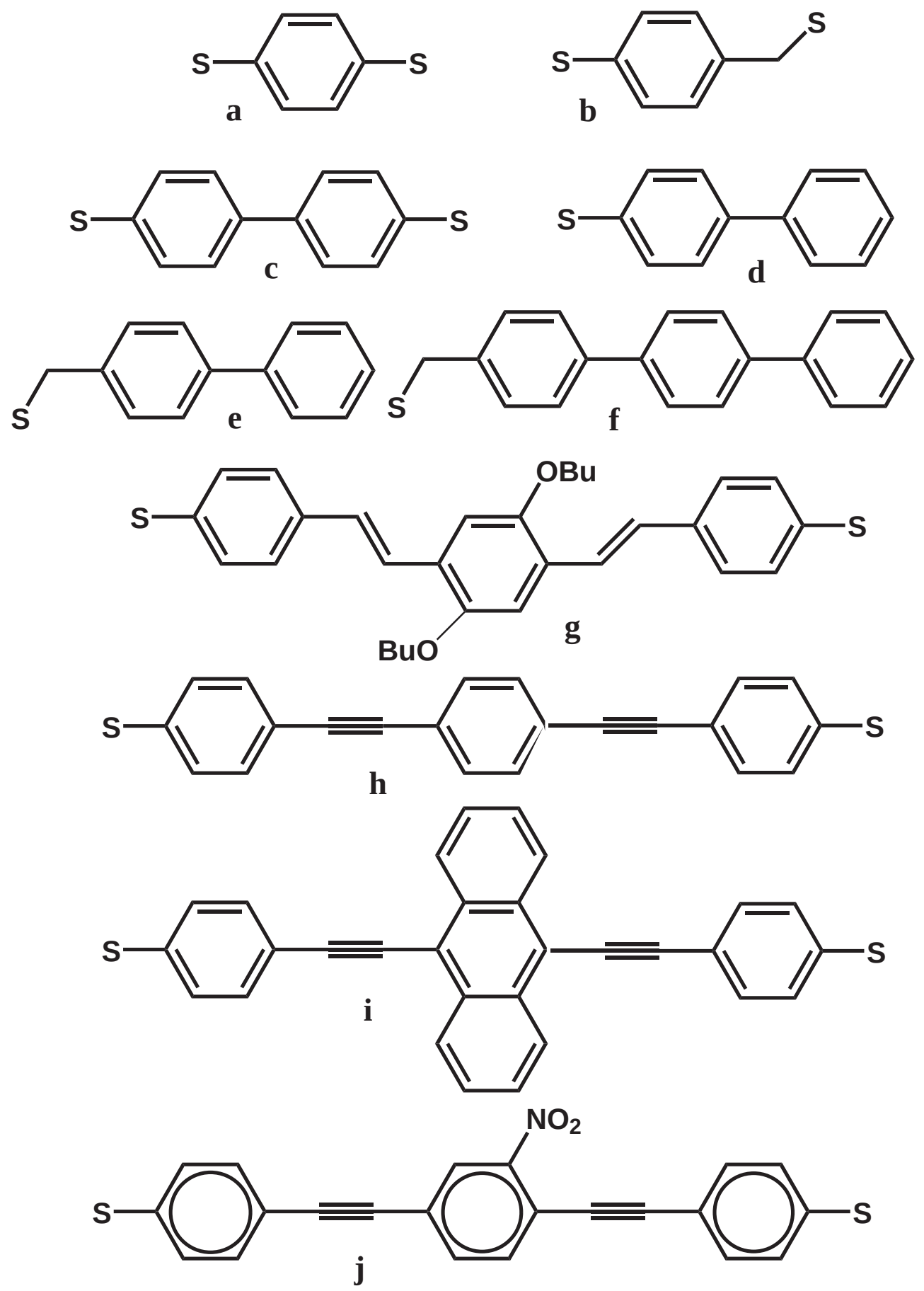

**Figure 18.** Chemical structure of the molecules listed in table 12.

Fundamental investigations of the correlation between molecular structure/conformation and electronic conductance of single molecules are crucial for deep understanding of the charge transport process, device operating mechanisms and further development of molecular electronics. Molecules based on oligothiophenes, phyenylacetylenes and phenylvinylnes have drawn much research interest owing to their good charge transport and chemical properties than that of other thiolated materials. Moth-Poulsen et al [73] systematically studied a series of single thiol end-capped oligo-phenylenevinylenes (OPVs) molecules by using a STM plus n-alkanethiol matrix technique. According the theoretical model described in a previous section, they calculated the decay constant β values for those molecules. Their results revealed that a big change in the electronic transparency of the various OPV derivatives due to the insertion of a methylene spacer group or nitro substituent. However, changes in the conjugation path through the central benzene ring from Para to Meta substitution do not affect the molecular electronic transmission too much. The experimental data is summarized in Table 13. Figure 19 shows the corresponding molecular structures.

**Table 13.** Comparison of molecular length, apparent height and decay constant β values for a series of molecular materials experimentally measured by using scanning probe microscopy techniques. Molecular structures are given in figure 19 [37,73-76]. [Adapted with permission from Ref. 73. Copyright (2005), American Chemical Society.]

| Entry | Molecule | Molecule length (Å) | Apparent height (Å) | β (Å) | Approach |
|---|---|---|---|---|---|
| 1 | Oligophenylene | N/A | N/A | 0.42±0.07 | CAFM (~2 nN) |
| 2 | **a** | 13.9 | 4.0 ±0.7 | 0.53±0.12 | STM |
| 3 | **b** | 14.2 | 3.5±1.0 | 0.65±0.16 | STM |
| 4 | **c** | 19.5 | 7.5±1.1 | 0.63±0.13 | STM |
| 5 | **d** | 19.4 | 6.1±2.0 | 0.78±0.23 | STM |
| 6 | **e** | 16.5 | 4.1±1.6 | 0.80±0.22 | STM |
| 7 | **f** | 15.9 | 4.2±0.9 | 0.72±0.13 | STM |
| 8 | **g** | 13.7 | 2.0±0.9 | 0.84±0.15 | STM |
| 9 | **h** | 19.3 | 4.70±1.01 | 0.94±0.12 | STM |
| 10 | **i** | 19.3 | 4.30±0.54 | 0.99±0.06 | STM |
| 11 | **j** | 14 | 0.0 | 1.2 | STM |
| 12 | **k** | 12 | N/A | 1.1 | CAFM (2 nN) |
| 13 | n-alkanedithiols | N/A | N/A | 0.53±0.03 | CAFM (2 nN) |
| 14 | n-alkanethiols | N/A | N/A | 0.94±0.6 | CAFM (2 nN) |
| 15 | n-$C_4SMe$ | N/A | N/A | 0.86±0.03 | STM |
| 16 | n-$C_4NH_2$ | N/A | N/A | 0.88±0.02 | STM |
| 17 | n-$C_4PMe_2$ | N/A | N/A | 0.97±0.02 | STM |

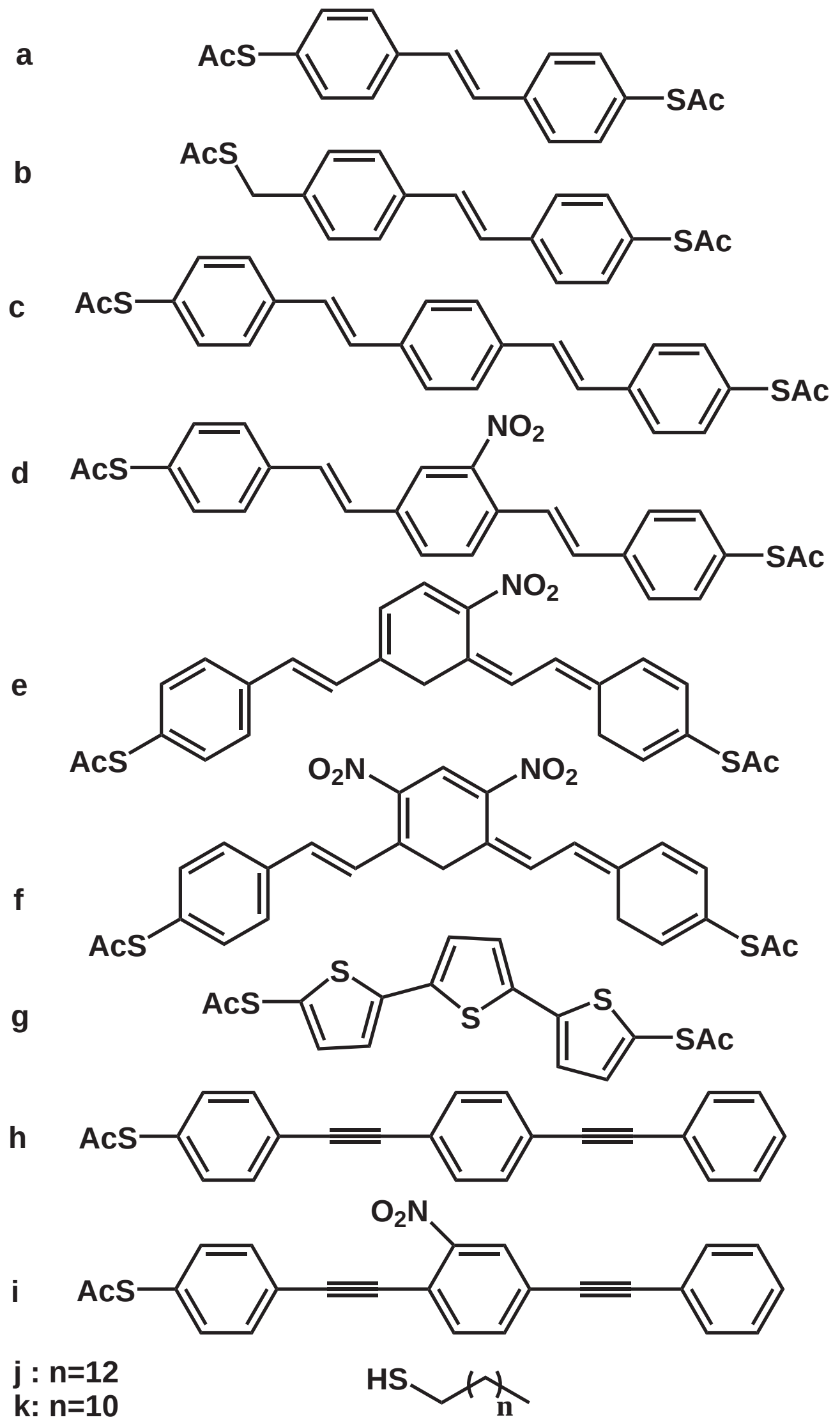

**Figure 19.** Chemical structures for the molecules studied by using STM shown in table 13.

Furthermore, other studies confirmed that the molecular conformation variation may affect the conductance of molecular junction to large extent. For example, Venkatarman et al [77] fabricated single molecular junction by using approach of breaking Au point contacts in the target molecular solution. They systematically investigated a series of pi-conjugated biphenyl molecular systems (see figure 20). As shown in Table 14, it was found that the molecular twist angle, altered by different aromatic ring substituents, can intensively affect the molecular conductance. The larger twist angle is, the lower of the molecular conductance.

**Table 14.** Effect of molecular conformation on the electronic conductance (unit in $G_0$). The single molecular junction is fabricated by using the method of breaking Au point-contacts in the corresponding molecular solution. [Adapted with permission from Ref. 77. Copyright (2006), Nature Publishing Group.]

| Molecule | Measured ($G_0$) | Calculated ($G_0$) | Peak width | Twist angle (º) |
|---|---|---|---|---|
| 1 | $6.4\times10^{-3}$ | $6.4\times10^{-3}$ | 0.4 | N/A |
| 2 | $1.54\times10^{-3}$ | $2.1\times10^{-3}$ | 0.8 | 0 |
| 3 | $1.37\times10^{-3}$ | $2.2\times10^{-3}$ | 0.8 | 17 |
| 4 | $1.16\times10^{-3}$ | $1.6\times10^{-3}$ | 0.9 | 34 |
| 5 | $6.5\times10^{-4}$ | $1.2\times10^{-3}$ | 1.3 | 48 |
| 6 | $4.9\times10^{-4}$ | $7.1\times10^{-4}$ | 0.6 | 52 |
| 7 | $3.7\times10^{-4}$ | $5.8\times10^{-4}$ | 0.9 | 62 |
| 8 | $7.6\times10^{-5}$ | $6.4\times10^{-5}$ | NA | 88 |
| 9 | $1.8\times10^{-4}$ | $3.5\times10^{-4}$ | 2.1 | N/A |

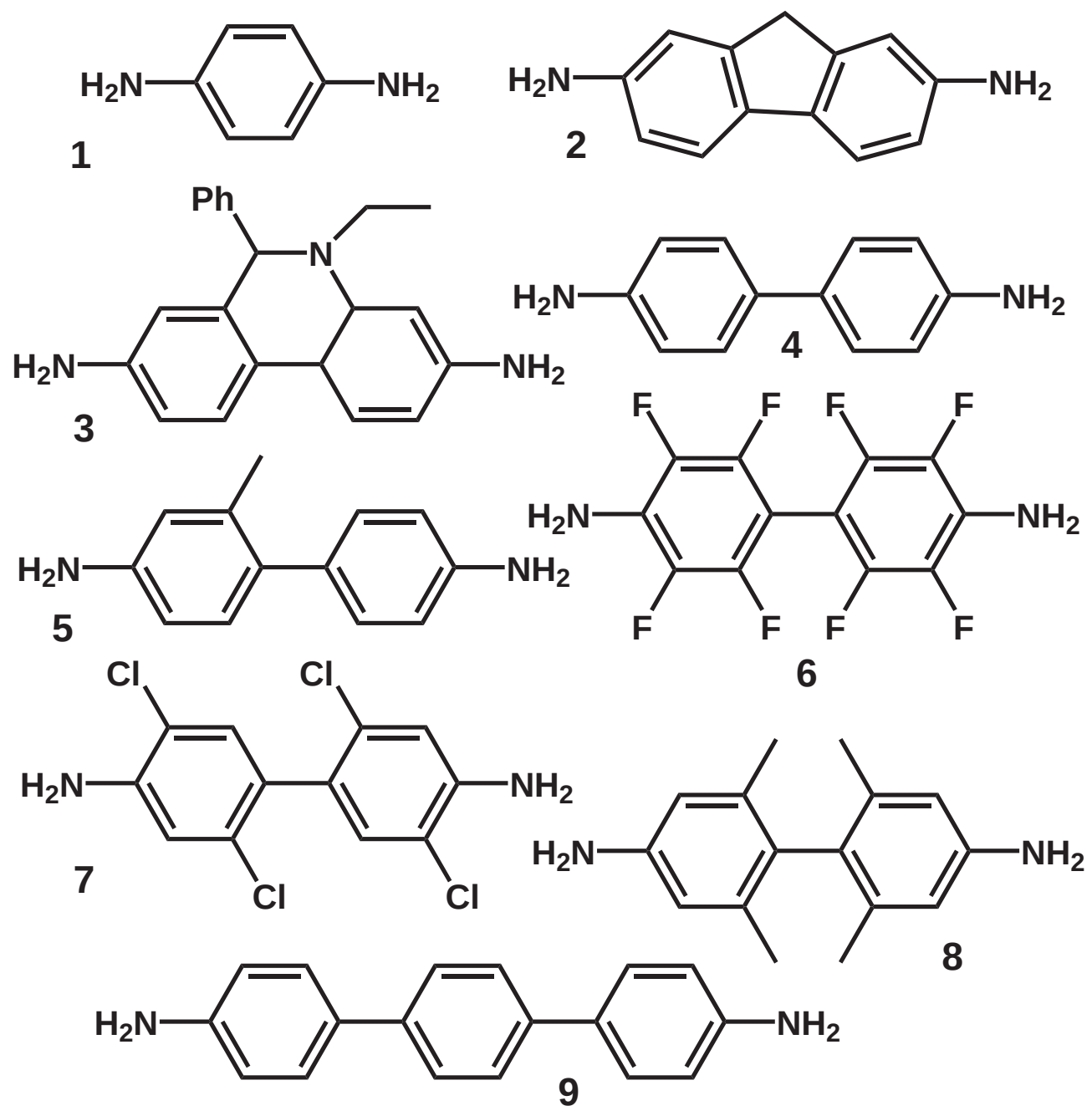

**Figure 20.** Chemical structure of the molecules listed in table 14.

Recently, Venkatarman et al. reported their experimental observations about the influence of molecular chemical substituents on its single-molecule junction conductance [35]. They use the same method as mentioned above to measure the low bias conductance of a series of substituted benzene diamine molecules in the molecular solution. In these molecular junction, the electron transport follows the mechanism of nonresonant tunneling, with the molecular conductance depending on the alignment of the electrode Fermi level to the closet molecular energy level. The author found that electron-donating substituents, which attempt to drive the highest occupied molecular orbital up. This will in turn result in higher molecular junction conductance. On the other hand, the electron-accepting functional groups pose the opposite effect. Their experiments revealed that the diamine molecules' HOMO is closet to the work function of gold electrode, confirming that the tunneling through these molecules is analogous to hole tunneling through an insulating film with varied potential barrier (depending on nature of the substituents). The data is listed in Table 15.

**Table 15.** Effect of molecular substituent on the molecular conductance. The ionization potential (IP) and relative conductance are also calculated and given out. [Adapted with permission from Ref. 35. Copyright (2007), American Chemical Society.]

| **Entry** | **Molecule name** | **Substituent (Numbers)** | **Calculated IP (eV)** | **Conductance peak ($10^{-3}$ $G_0$)** | **Calculated relative conductance ($10^{-3}$ $G_0$)** |
|---|---|---|---|---|---|
| 1 | Tetramethyl-1,4-diaminobenzene | $CH_3$ (×4) | 6.36 | 8.2 × 0.2 | 7.6 |
| 2 | 2,5-dimethyl-1,4-diaminobenzene | $CH_3$ (×2) | 6.59 | 6.9 × 0.2 | 6.7 |
| 3 | 2-methoxy-1,4-diaminobenzene | $OCH_3$ (×1) | 6.55 | 6.9 × 0.2 | 7.1 |
| 4 | 2-methyl-1,4-diaminobenzene | $CH_3$ (×1) | 6.72 | 6.4 × 0.6 | 6.5 |
| 5 | 1,4-diaminobenzene | H (×4) | 6.83 | 6.4 × 0.2 | 6.4 |
| 6 | 2,5-dichloro-1,4 diaminobenzene | Cl (×2) | 7.14 | 6.1 × 0.2 | 6.0 |
| 7 | 2-bromo-1,4-diaminobenzene | Br (×1) | 7.02 | 6.1 × 0.6 | 6.1 |
| 8 | Trifluoromethyl-1,4-diaminobenzene | $CF_3$ (×1) | 7.22 | 6.1 × 0.2 | 6.2 |
| 9 | 2-chloro-1,4-diaminobenzene | Cl (×1) | 7.00 | 6.0 × 0.4 | 6.2 |
| 10 | 2-cyano-1,4-diaminobenzene | CN (×1) | 7.30 | 6.0 × 0.3 | 5.9 |
| 11 | 2-fluoro-1,4-diaminobenzene | F (×1) | 7.03 | 5.8 × 0.4 | 6.3 |
| 12 | Tetrafluoro-1,4-diaminobenzene | F (×4) | 7.56 | 5.5 × 0.3 | 5.2 |

Theoretical calculation results also concluded that the I-V shape of a molecular junction can be largely determined by the electronic structure of the molecule itself, while the presence of electrode/molecule interfaces play a key role in determining the absolute value of the device current (or in another word the device conductance) [78]. For example, Majumder et al. carried out theoretical calculations on the effects of molecular structure and binding metal atoms on the electronic structure of a series of conjugated molecules. As shown in Table 16 [79], they performed a first-principle electronic structure calculations using Hartree-Fock (HF), density functional theory (DFT) and HF-DFT hybrid methods for a series of molecular wires (shown in Figure 21). The calculated energy gaps for molecules (**a** to **g**) are 4.52, 4.14, 3.79, 3.37, 2.27, 2.56 and 2.84, respectively. The results indicated that about 2.2 eV bias voltage is required to achieve electron transport through the target molecule. The connection with metal atoms (Au, Ag and Cu) will lead to an obvious reduction in the molecular energy gap compared to that of the free molecules.

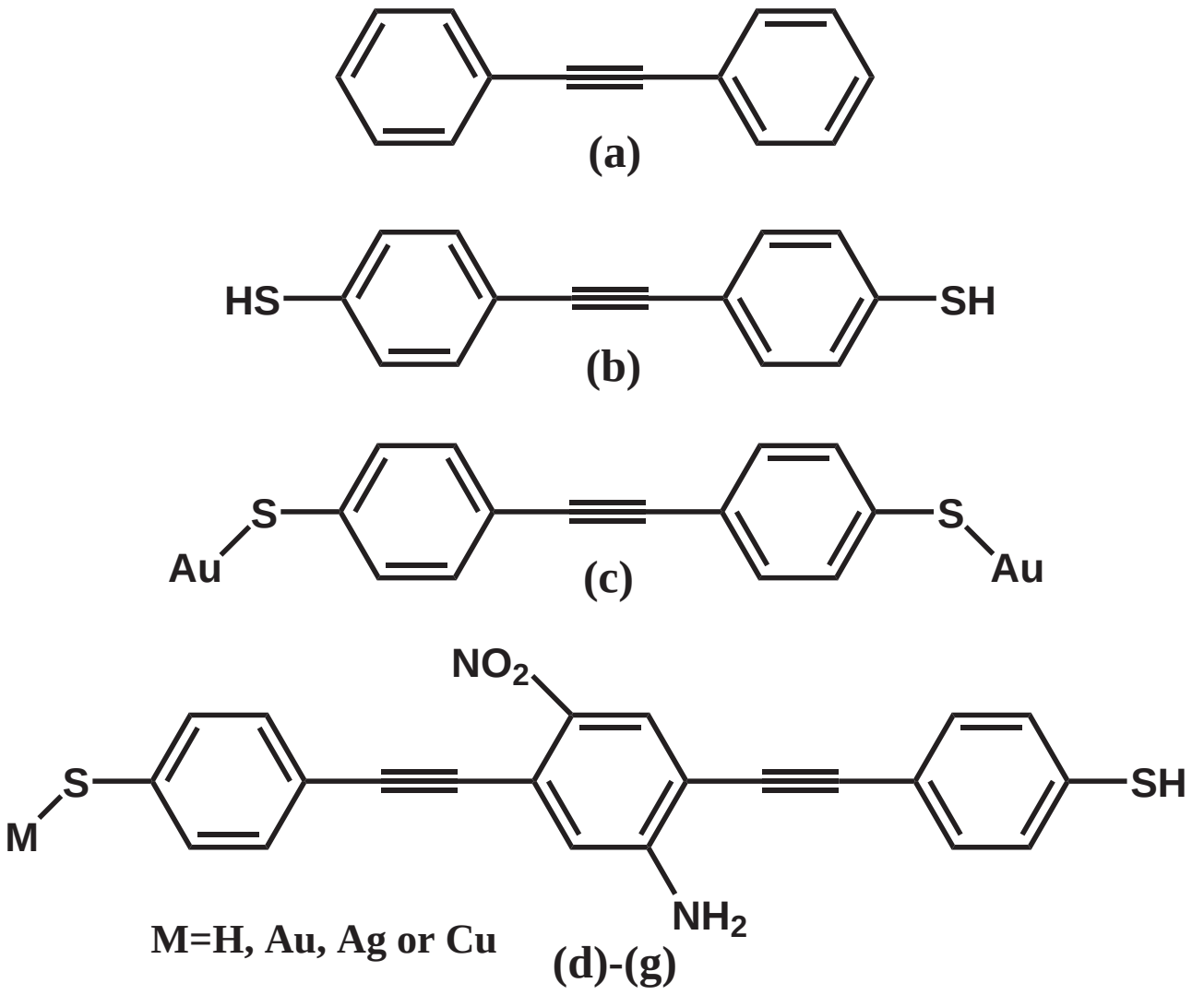

**Figure 21.** Molecular structures of a series of conjugated molecules calculated.

**Table 16.** First-principle molecular electronic structure calculated using Hartree-Fock (HF), density functional theory (DFT) and HF-DFT hybrid methods. [Adapted with permission from Ref. 79. Copyright (2002), The Japan Society of Applied Physics.]

| Molecule | HOMO (eV) | LUMO (eV) | Energy gap (eV) |
|---|---|---|---|
| (a) | -6.00 | -1.48 | 4.52 |
| (b) | -5.97 | -1.65 | 4.14 |
| (c) | -6.13 | -2.34 | 3.79 |
| (d) M=H | -6.25 | -2.88 | -3.37 |
| (e) M=Au | -6.00 | -3.729 | -2.27 |
| (f) M=Ag | -5.79 | -3.23 | -2.56 |
| (g) M=Cu | -5.81 | -2.79 | -2.84 |

### *7.4.1.3 Organosilane-based Self-Assembly Monolayers*

In section 2 of this chapter, we introduced that organosilane-based self-assembly monolayer is a very useful active media in molecular electronics. Collet et al investigated the formation and modification of a series of organosilane monolayers through measuring their electrical properties in junction form of Al/Silane SAM/Si [9]. They firstly fabricated a vinyl-terminated trichlorosilane monolayer on a n type Si (100) following the procedures descried in section 2 of this chapter. Once the monolayer is grafted, they can obtain –$CH_2OH$ or –COOH terminated monolayers through chemical reactions of hydroboration, hydrolysis, or oxidation. Furthermore, they can attach highly conjugated moieties onto these SAMs by using esterification reactions between the –COOH end-groups and different alcohols (such as benzyl alcohol, retinal, pyrene methanol). The resulted SAM is schematically shown in figure 22. The authors performed physical characterizations by measuring the monolayer embedded in a Al/SAM/Si junctions. Their results are summarized in table 17. It is indicated that the good insulating properties of the SAM are not altered by the chemical modification process, while the permittivity of the monolayer can be modified via adding different dipolar moieties. From the table, we can see that the silane monolayers with phenyl and pyrene ending groups have a lower barrier height and higher current density at the same bias voltage than that of the others. This is due to the conjugation effect of the monolayers. Again, the experimental results remind us that the molecular chemical structure and functional groups have a large effect on the device physical performances.

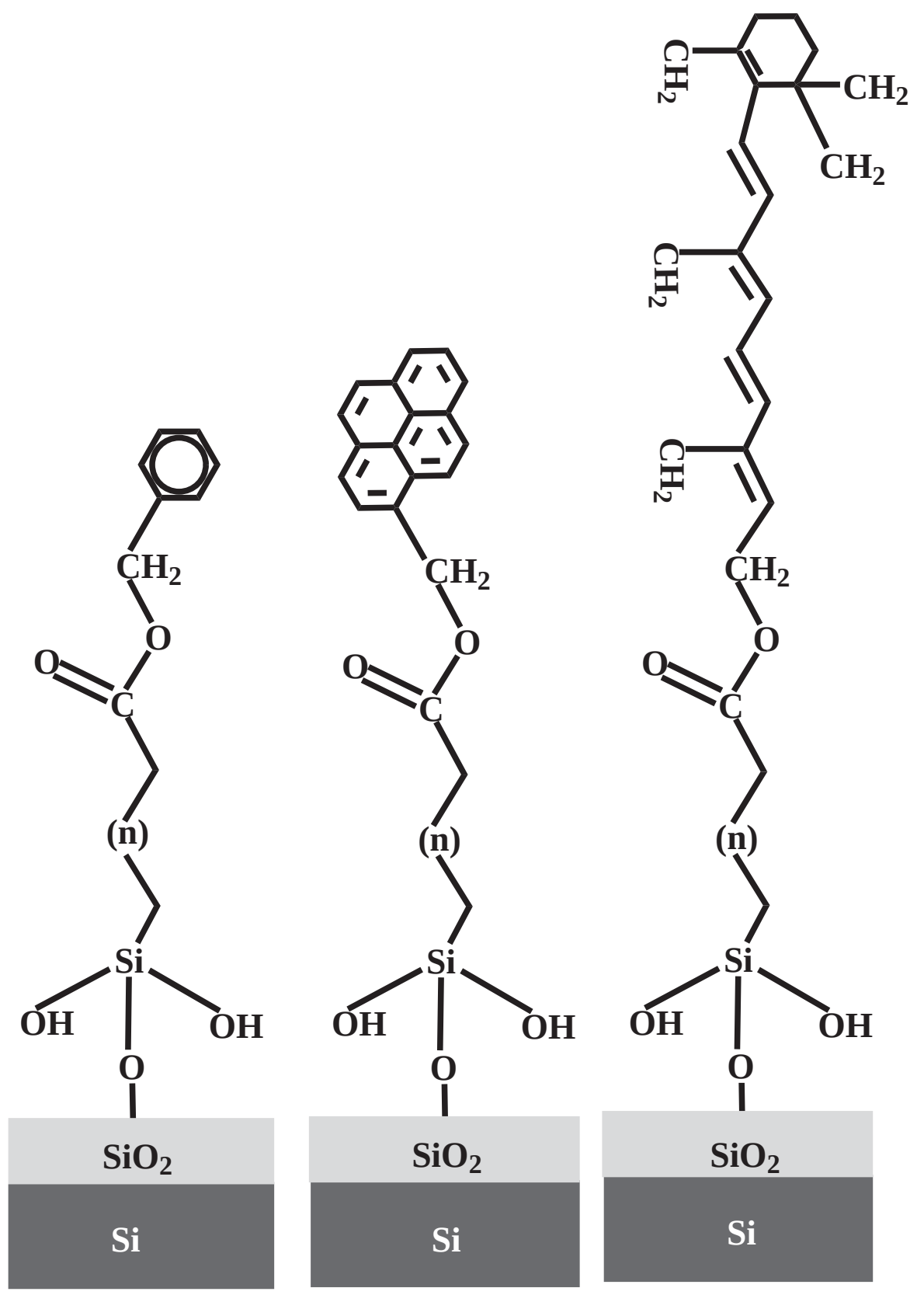

**Figure 22.** Organosilane monolayer can be used as surface anchor to direct the growth of a variety of conjugated parts on the substrate (from left to right: phenyl, pyrene and retinal ended monolayer). This technique will be useful in the bottom-up approach of molecular memory system. [Adapted with permission from Ref. 9. Copyright (1997), Elsevier B.V.]

**Table 17.** Summary of electrical properties of organosilane monolayers with different ending groups with junction structure of Al/Silane SAM/Si. The substrate is n type Si (100). The values in parenthesis are calculated ones, while n is the optical indice of the monolayer. [Adapted with permission from Ref. 9. Copyright (1997), Elsevier B.V.]

| **Ending Group** | **Current at 1V (A/cm$^2$)** | **Barrier Height (eV)** | **Monolayer Capacitance (μ F/cm$^2$)** | **Ellipsometry n.d (nm)** | **Monolayer permittivity** | **Monolayer thickness (nm)** |
|---|---|---|---|---|---|---|
| -$CH_3$ | $10^{-8}$ | 4.3-4.5 | 1.07 | 2.75 | 2.2 | 1.84 (1.86) |
| -CH=$CH_2$ | 3x$10^{-8}$ | 4.1-4.2 | 0.78 | 2.81 | 1.82 | 2.1 (2.1) |
| -COOH | 8x$10^{-8}$ | 4.1-4.2 | 0.95 | 3.79 | 2.55 | 2.4-(2.1) |
| -$CH_2$OH | 2x$10^{-8}$ | 4.1-4.2 | N/A | N/A | N/A | N/A |
| -COO$CH_2$-Phenyl | 5x$10^{-8}$ | 3.6-3.8 | 0.85 | 5.03 | 2.85 | 2.98 (2.73) |
| -COO-Retinol | 2x$10^{-8}$ | N/A | 0.75 | 6.12 | 3 | 3.53 (3.48) |
| -COO$CH_2$-Pyrene | $10^{-7}$ | 3.9 | 0.89 | 4.92 | 2.9 | 2.9 (2.98) |

The silane monolayers can be further used in patterning or directing placement of various nano-materials like nanoparticles and nanowires. Ma et al reported their investigations on the electrostatic funneling for precise nanoparticle placement [11]. They use the current CMOS (complementary metal-oxide-semiconductor transistor) fabrication technology to define a pattern of structures on the substrate, which can be then selectively modified by using different SAMs (see figure 23 (a) and (b)). The $-NH_3$ terminated silane molecules will be formed on silicon dioxide surfaces, while the $-COO^-$ terminated molecules will be selectively grown on gold surfaces. In this way, they can obtain a positively and negatively charged patterns on the substrate in an aqueous solution. When such a substrate is immersed into a colloidal solution containing negatively charged Au nanoparticles, the nanoparticles will be directed into the positively charge areas. Using a similar approach, Morrill et al recently realized the selective placement of silver nanoparticles on surfaces covered with 3-aminopropyltriethoxysilane SAM. The mechanism arises from the silver's ability to donate electrons to the nitrogen's anti-bonding orbital via π back bonding (see figure 23(c)). These knowledge and techniques can be further used in fabrication of molecular junctions with specially designed functional active units and metal/molecule interfaces that will enable us the molecular-level ability to engineer the molecular device.

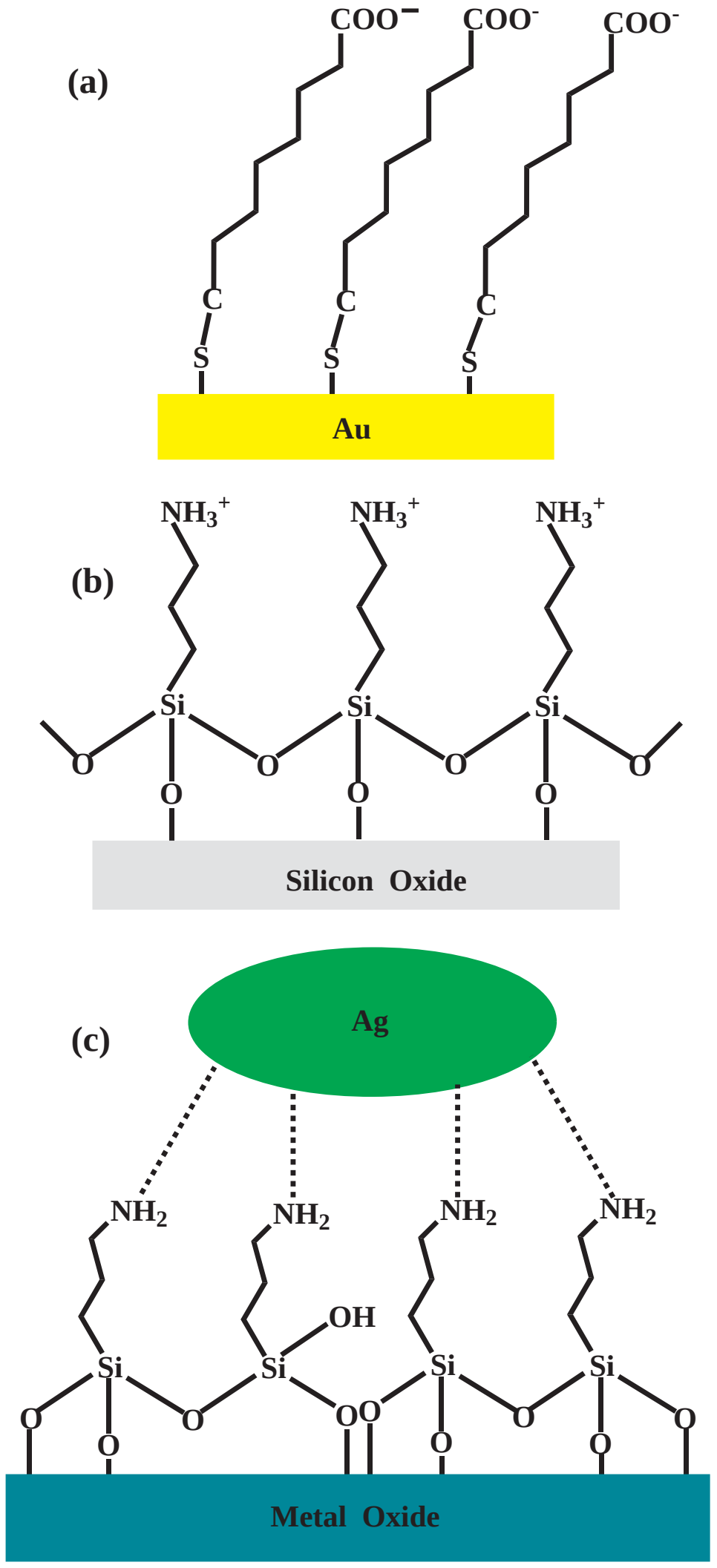

**Figure 23.** Organosilane monolayers can be used to modify the substrate surface and direct the assembly of metal nanoparticles. (a) $-COO^-$ terminated SAM is selectively formed on gold surfaces and (b) $-NH_3^+$ terminated molecules are selectively formed onto silicon dioxide surfaces [11]. (c) Silver nanoparticle is selectively attached to the $-NH_2$ terminated aminosilane surfaces [12].

## 7.4.2 Molecular Switching and Memory

One big obstacle in studying the device performance of SAMs is the uncertainty of the molecule numbers within the molecular junction area. Generally, it is not known exactly how many molecules contribute to the device's conduction. In addition, we are not very sure if there are any intermolecular effects between the neighboring molecules. For molecular scale information memory, random fluctuations resulted from intermolecular interactions may have a strong influence on the junction electrical/optical properties. All these factors will make it complicated both for theoretical understanding efforts and for experimental investigations.

### *7.4.2.1 Patterning Self-Assembly Monolayers*

To circumvent the problems encountered in monolayers molecular electronics, one simple way is to get average value for each data point based on repeatedly measuring thousands of molecular junctions.

Molecular matrix method is another effective yet simple way to be employed in molecular electronics [75,80,81]. In this approach, the target molecules are embedded in relatively insulate molecular matrix (usually made from long chain n-alkanethiol such as 1-Dodecanethiol). We just co-dissolve the target molecular materials and matrix molecules in the same solution with an appropriate molar ratio (e.g. 5 : 100 = target molecule: 1-dodecanethiol). This will result in the formation of a mixed SAM with the target molecules surrounded by more insulated n-alkanethiols. Figure 24 presents an example for this approach, where a conjugated molecule is inserted into Dodecanethiol monolayer.

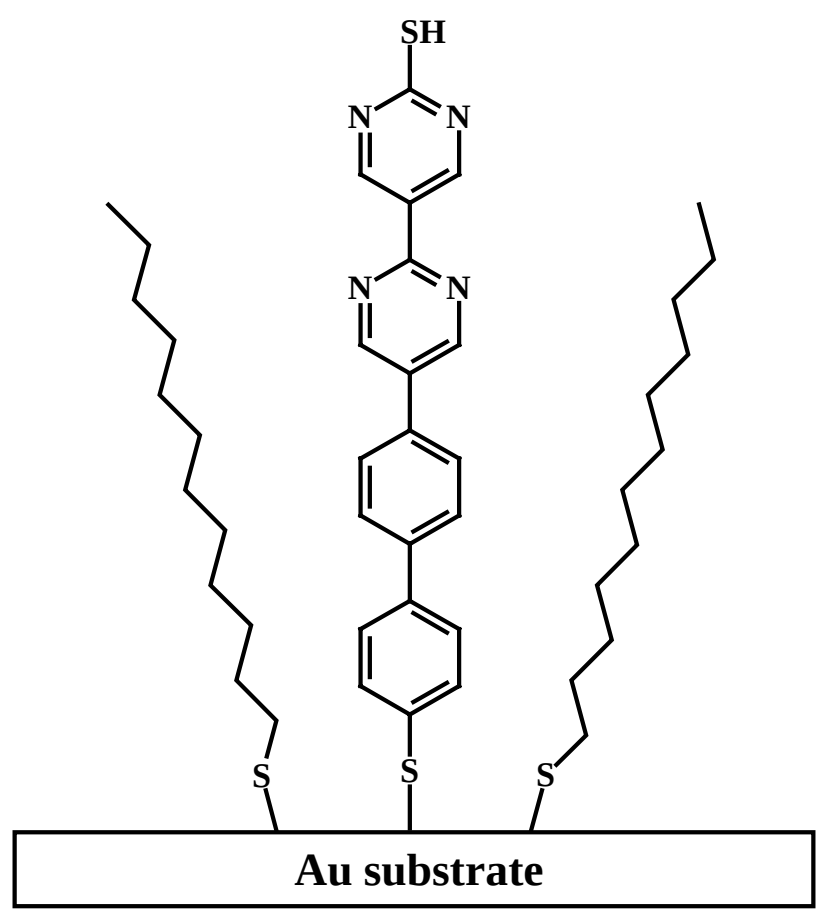

**Figure 24.** Schematic structure showing a conjugated dithiol molecular wire inserted within 1-dodecanethiol matrix.

As shown figure 25, there are some other techniques to fabricated SAM's patterns on various substrates at different conditions. It may includes: 1) Stamp-printing [82]; 2) selective growth on pre-patterned substrate [6]; 3) AFM tip-based dip-pen writing or lithography [83-85] and 4) STM tip-based lithography [86]. Table 18 summarized and compared these methods. It is indicated that SPM-based methods provide a higher resolution and reliable results, which is promising for nanoscale information memory applications. On the other hand, the lithography approaches usually have a poor resolution and involve steps of using photoresist and organic solvents. The lithography methods are obviously unfavorable for using in molecular electronics.

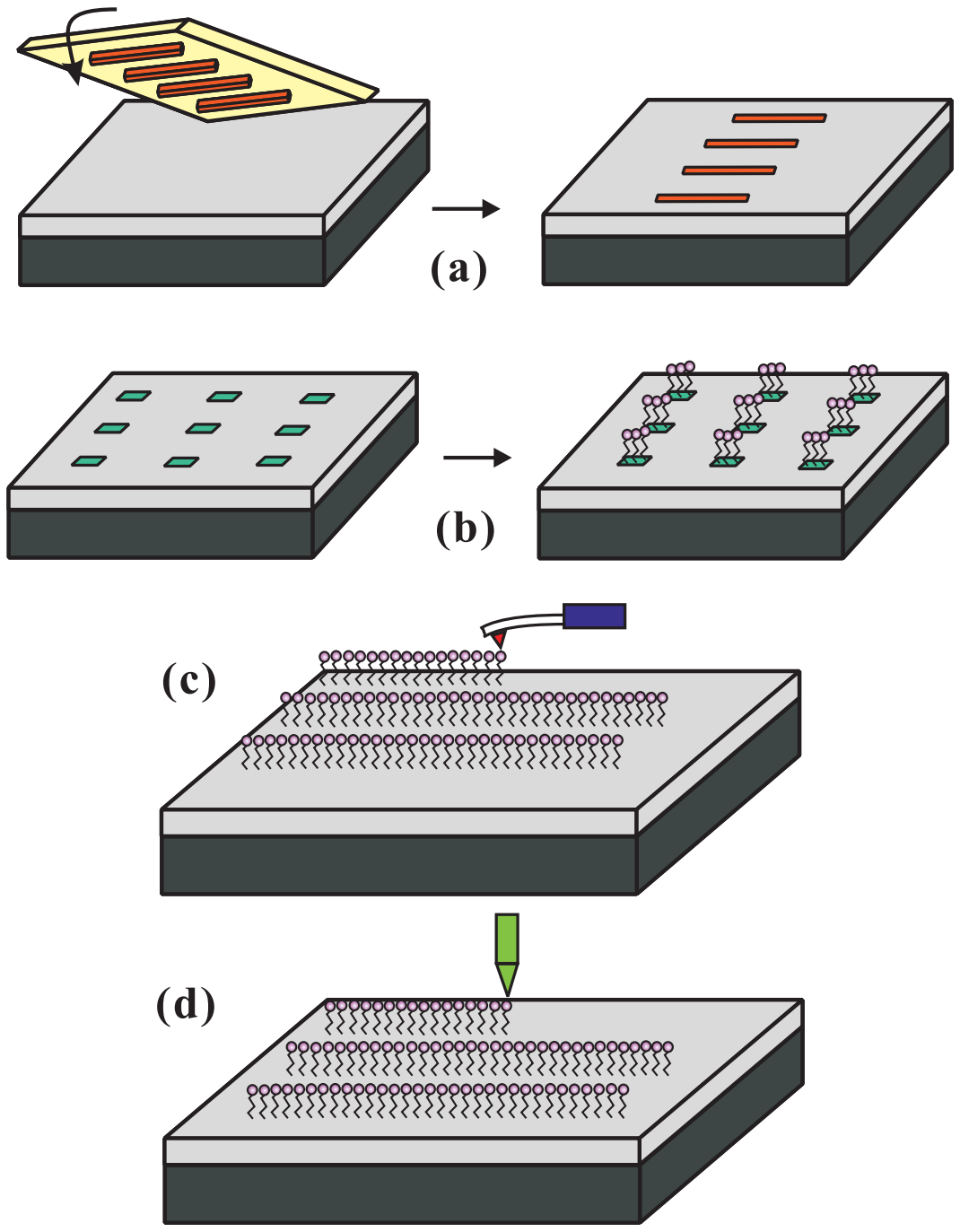

**Figure 25.** Schematic process for patterning SAM on substrate: (a) Stamp-printing, (b) selective growth on pre-patterned substrate (e.g. lithography pre-defined Au regions), (c) AFM tip-based dip-pen writing and (d) STM tip-based lithography.

**Table 18.** Comparison of the fabrication approaches for self-assembly monolayer patterning.

| Approach | Process | Resolution (nm) | Advantages | Disadvantage | Refs. |
|---|---|---|---|---|---|
| Photolithography | Stamp printing | 5000 | Low cost and mass production | Poor resolution, contamination from stamp | [82,87] |
| | Selective growth on pre-patterned substrate | 2000 | Low cost | Low resolution | [6] |
| E-beam lithography | Stamp printing | 200 | medium resolution and cost | High-cost, contamination from stamp | N/A |
| | Selective growth on pre-patterned substrate | 50 | Good resolution and compatibility with semiconductor industry | High-cost, complicated process | N/A |
| Atomic force microscopy | Dip-pen writing | 100 | Good resolution | reproducible | [84,88] |
| | CAFM tip-based lithography | 50 | High resolution | Meddle-cost, reliable | [83,85,89,90,] |
| Scanning tunneling microscopy | STM tip-based lithography | 2 | Highest resolution | High-cost, low-reliability | [86] |

### 7.4.2.2 Switching of Self-Assembly Monolayer Devices

Low-temperature characterization under high vacuum conditions is a powerful way to eliminate most of the unfavorable factors for the junctions' electrical and optical properties. Referring to the ultra-high vacuum system given in section 2 of this chapter, various control experiments can be easily conducted with fine variation of the temperature, oxygen, water, solvent or optical illumination. Since the molecules are in their low energetically state, the measurements at low temperatures will offer a good chance to explore the molecular-specific properties and its effect on the device characteristics.

In a previous work [6], we reported the observation of high photo-responsivity with intrinsic amplification for a novel sandwich structure. The devices are made from molecular monolayers softly sandwiched between two thin gold electrodes. The bottom electrodes were fabricated by using photolithography method, while the top electrode bars were printed from elastic stamps made from a polymer. The experimental parameters and procedures for making polymer stamps are given in tables 19 and 20. The experiments are conducted under conditions of high vacuum and low temperature. As shown in figure 26, the junctions with a set of molecular wires, with different length and ending groups and conjugation degrees, are comparably investigated through current-voltage measurements at low temperatures. The device showed reversible optoelectronic switching with on/off ratio of 3 orders of magnitude at 95 K. The switching phenomenon is independent of both optical wavelength and molecular structure, while it strongly depends on the temperature. However, the switching on/off ratio is dependent on the molecular structure to some extent. The higher conjugation of the molecular wires is, the higher on/off ratio. The switching ratio for molecular junctions with dithiol molecules is relatively lower than that of the molecules with single thiolated group. The results are very intriguing, potentially providing a novel method to obtain extremely sensitive and fast, UV to IR, photodetectors.

**Table 19.** Silicon wafer preparation for photoresist lithography patterning.

| Step | Action | Time (s) |
|---|---|---|
| 1 | Nitrogen gun to blow clean the holder and close lid after loading the sample. | N/A |
| 2 | Ultrasound in 50 mL Acetone bath | 20 |
| 3 | Ultrasound in 50 mL Methanol bath | 20 |
| 4 | DI rinse under running water | 60 |
| 5 | Remove from beaker bottom and nitrogen gun blow dry | N/A |

**Table 20.** Experimental procedure for S1813 photoresist photolithography.

| Step | Action | Time (s) |
|---|---|---|
| 1 | Prepare silicon wafer according to RCA cleaning procedure | N/A |
| 2 | Center clean wafer on spinner, set to 4000 rpm | 45 |
| 3 | With clean pipette cover entire wafer with S1813 photoresist | N/A |
| 4 | Spin wafer at 4000 rpm | 45 |
| 5 | Soft-bake at 90C (keep covered with petri dish but allow ventilation) | 180 |
| 6 | Cooling down (keep covered with petri dish) | 60 |
| 7 | Blow off any junk that may be on mask | N/A |
| 8 | Check the wavelength and power of the UV light | N/A |
| 9 | Pattern exposure | 13 |
| 10 | Bake at 110 C (keep covered with petri dish) | 150 |
| 11 | Cooling down | 60 |
| 12 | Develop in 1:1 of photoresist Developer : DI water mixed solution | 75 |
| 13 | Rinse the sample with DI water flow and dry with nitrogen gun | N/A |

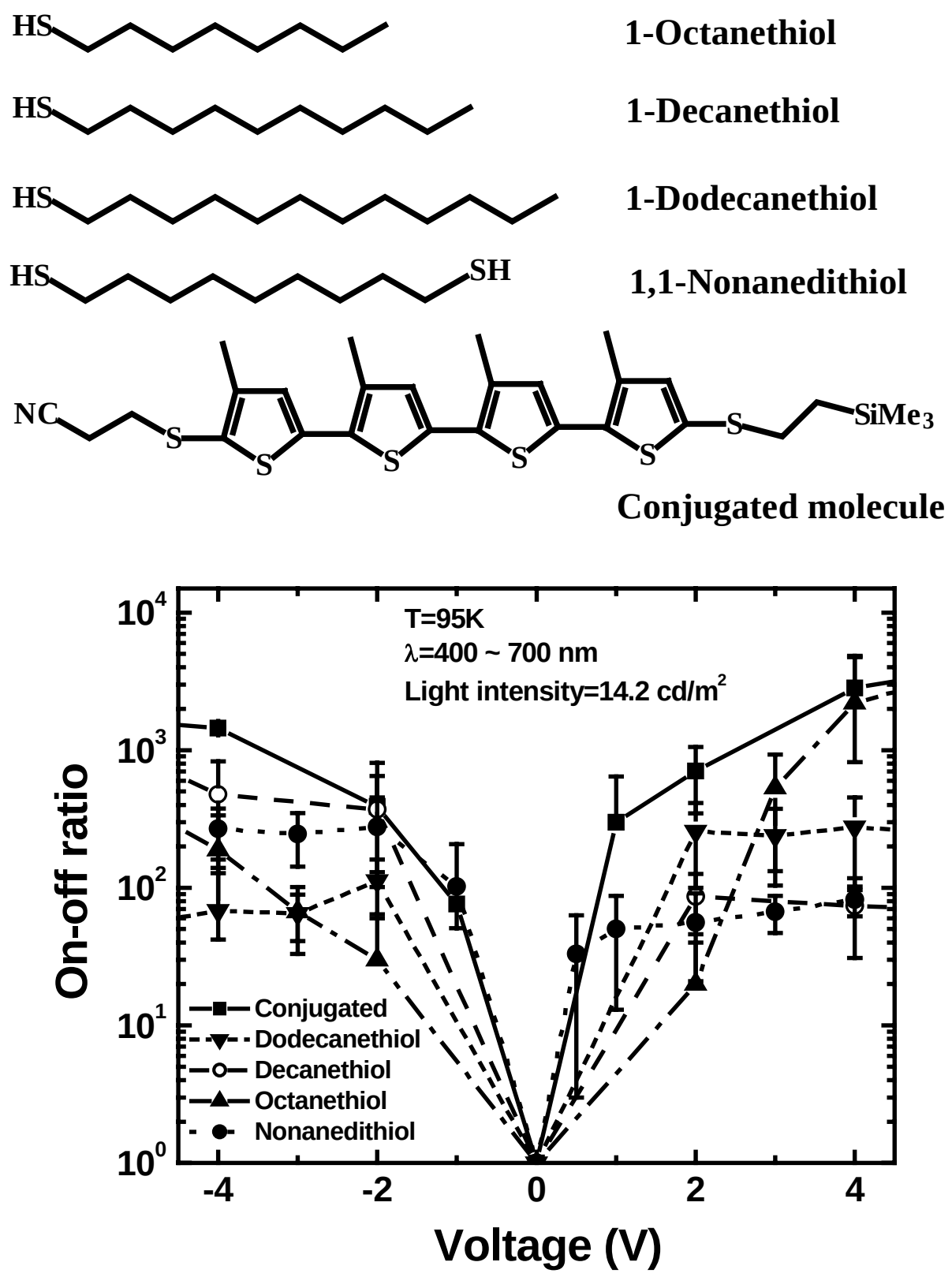

**Figure 26.** Chemical structure of molecular wires studied (top). Effect of molecular structure on the current switching ratio between light and dark conditions (bottom). It is clear that the junction with the conjugated molecules can yield a higher switching ratio than that without [6].

Temperature dependence results are shown in figure 27. It is indicated that the Au/molecule interfaces in the crossbar junctions have distinct chemical nature and different electron tunneling parameters. The charge transport in the junction can be mainly attributed to Richardson-Schottky thermionic emission process under light condition, while Fowler-Nordheim tunneling plays a dominant role in dark (refer to previous section for these models). An explanation for the optical switching is that the injected light induces some physical changes within the crossbar junction. It may be the tunneling parameter at the top wire/lead interface and/or the transport mechanism through the wire. Two evidences support this hypothesis. Firstly, under light irradiation, the temperature dependence of the junction is much weaker than that in dark, especially for the junctions with conjugated wires. This behavior is a clear indication of a change in the conduction mechanism, which may result from the interaction between the photons and the trapped electrons in the wires. Secondly, no similar optical switching can be detected from scanning tunneling microscopy measurements of the alkanethiols at room temperature, where there is no such weakly coupled metal/molecule contacts. Again, our work shows the importance of metal/molecule interface, molecular ending groups and conformations. For further applications of this kind of molecular devices, one of the big challenges is how to realize single molecule-scale molecular junctions with an ability of addressing the same junction at various temperature and optical illuminating conditions.

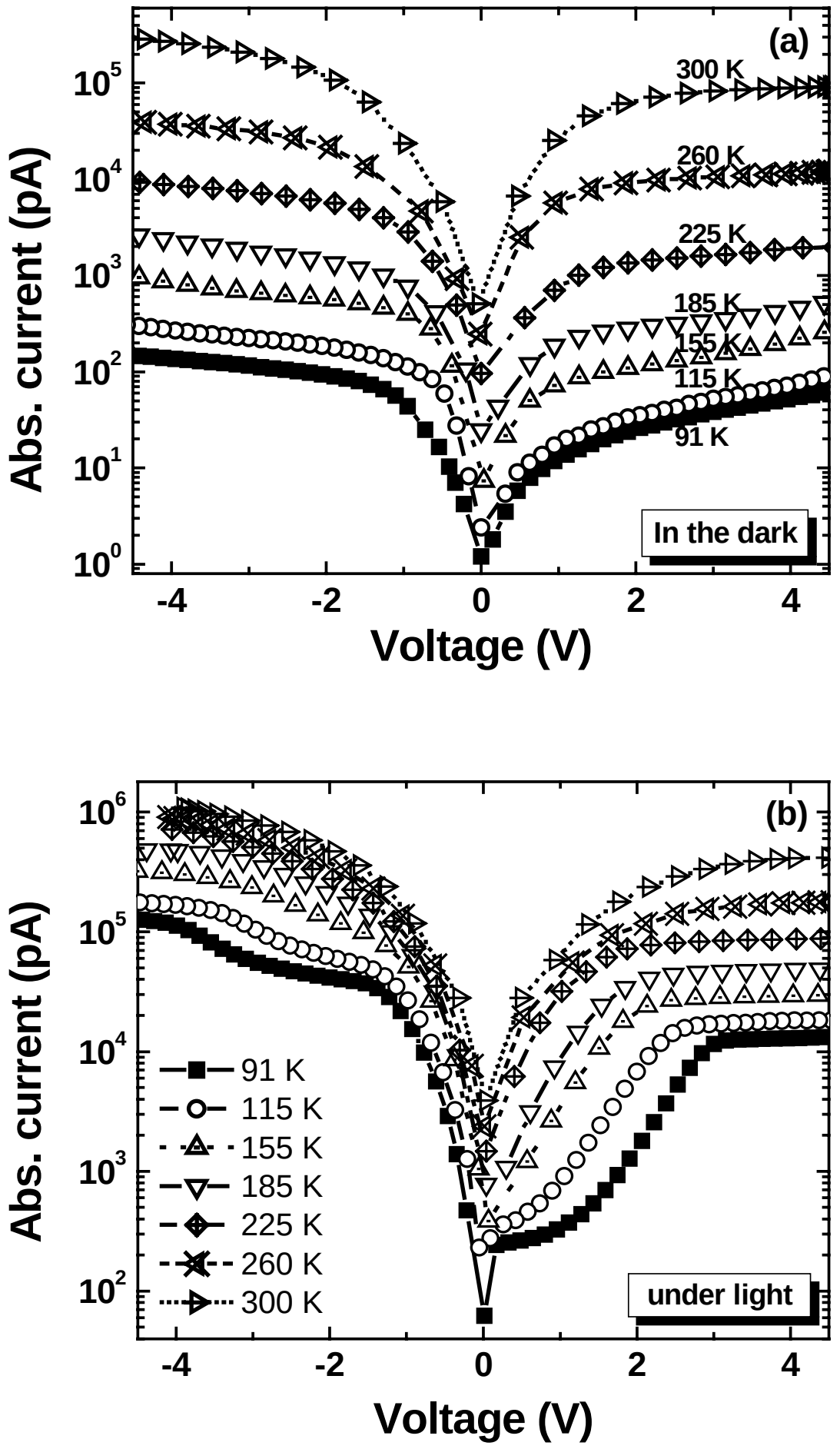

**Figure 27.** *I–V* temperature dependence of a 1-decanethiol junction (a) in the dark and (b) under illumination of florescent light. [Reprinted with permission from Ref. 6. Copyright (2009), Elsevier B.V.]

## 7.5 Other Organic Materials-based Memory

There are some other kinds of molecular functional materials that can be used as active layers in nanoscale data storage. Such organic materials may include polymers, charge transfer salts and Langmuir-Blodgett films. In this section, we just give a very brief introduction about the recent advances in the area of polymer-based molecular memory. For more information, the readers can refer to the good review articles cited and the references therein.

Functional polymers are the other kind of molecular materials for information memory applications. Switching of polymer thin film devices have been observed with mechanisms of charge trapping/detrapping, doping/dedoping and proton trapping/detrapping [91-94]. The widely studied polymers include polypyrrole, polythiophene, polyaniline, polybipyridinium, and PMMA [95,96]. The growth of polymer thin layer has been introduced in the second section of this chapter, which includes spin-coating, dip-coating, drop-casting and vacuum spray. The synthesis, fabrication, characterization of polymer memory materials and devices are too large to review in detail within the boundaries of this chapter. Here, our purpose is to give some examples about this widely studied material. Figure 28 presents the molecular structure of a group of polymers with electrical switching properties and their memory performances are summarized in table 21 [97]. It is shown that the switching time, on/off ratio and threshold voltage for polymer junctions are comparable to that of the devices made from small molecular weight redox materials and SAMs. However, it will be a great challenge to fabricate nanoscale polymer junctions due to the big difficulty in fabricating high quality polymer monolayers.

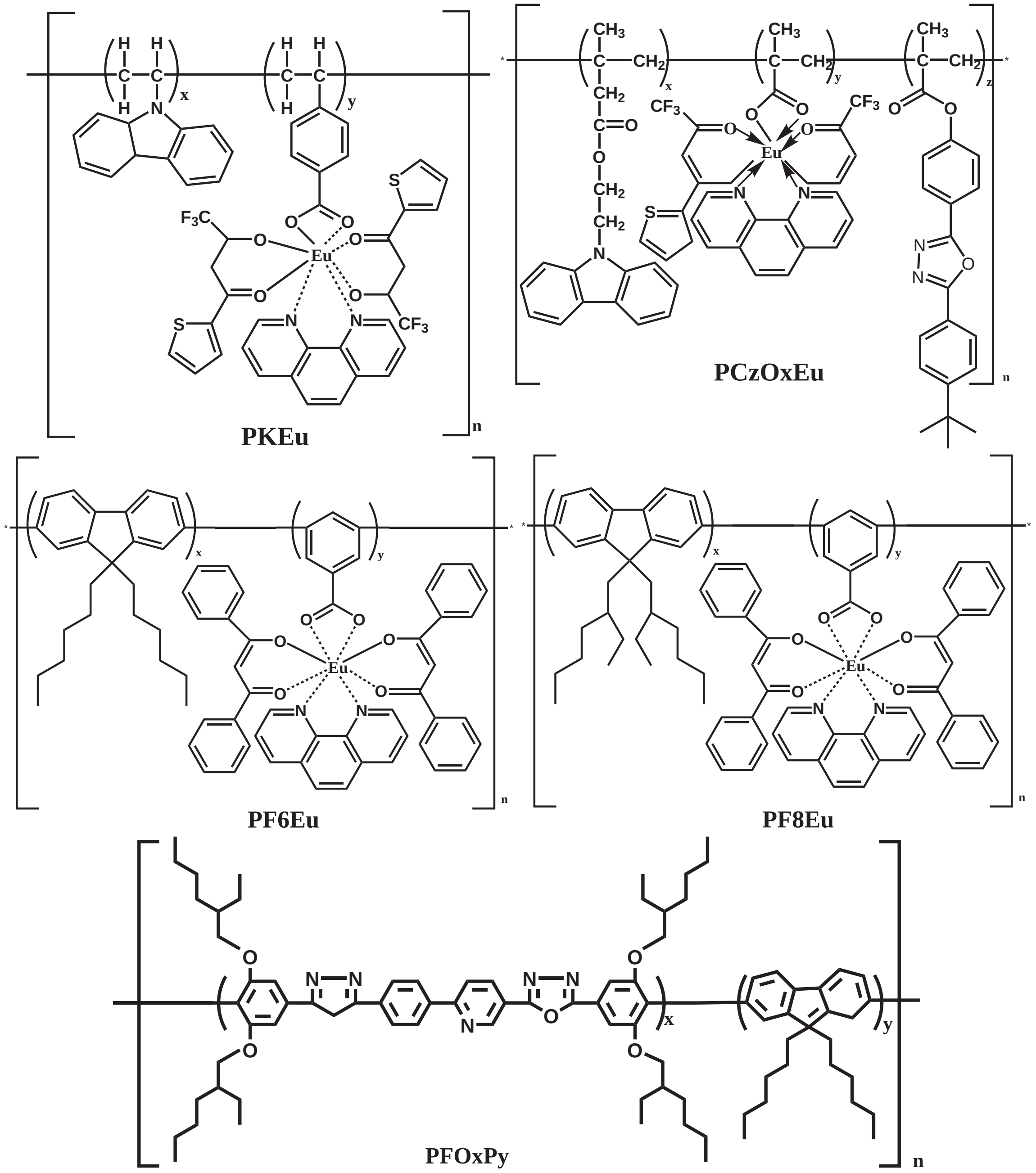

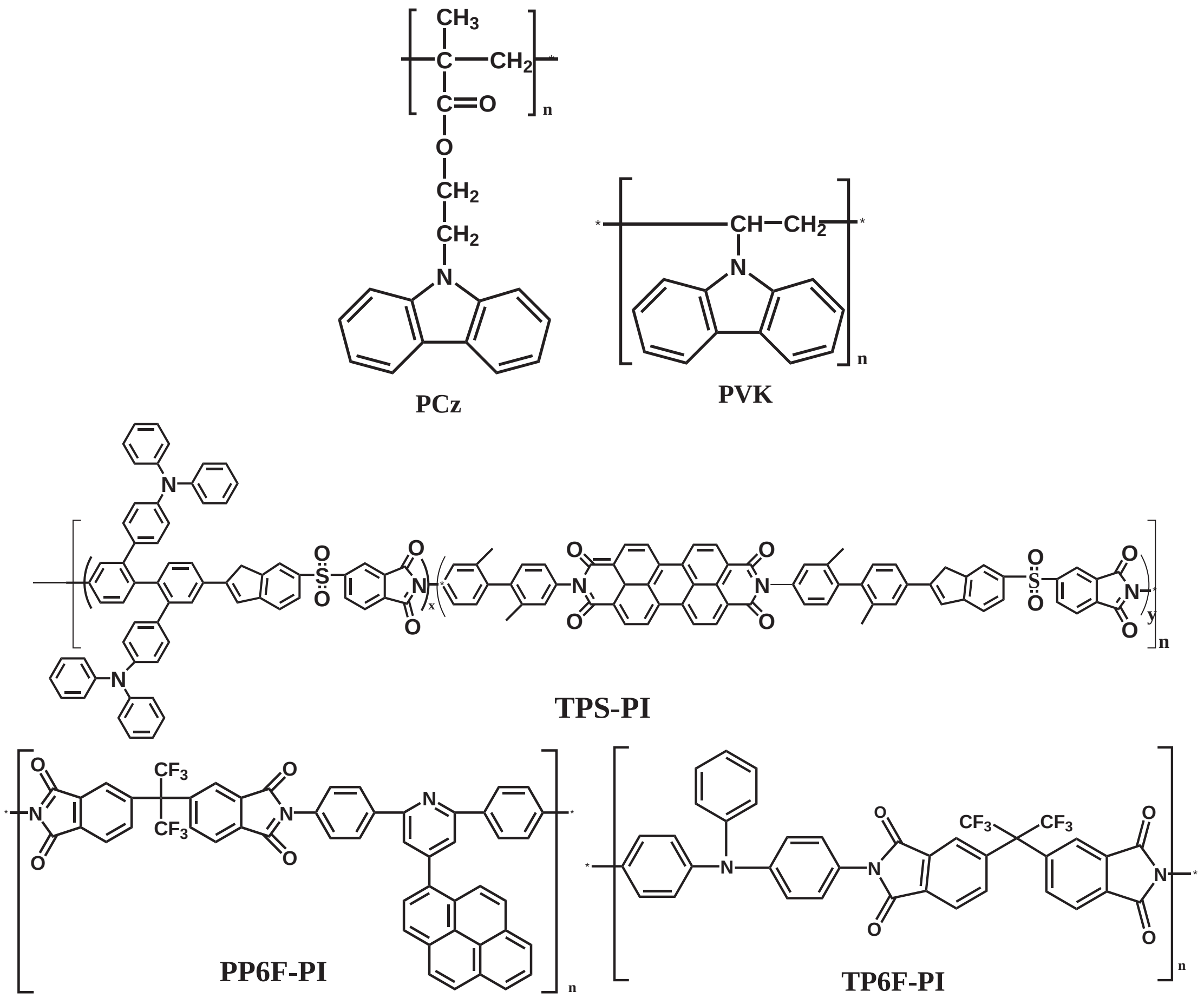

**Figure 28.** Molecular structure of the polymers with electrical switching properties [97].

**Table 21.** Summary of the device performance evaluation results of polymer memories. [Adapted with permission from Ref. 97. Copyright (2007), Elsevier B.V. ]

| Devices | Memory effects | Write voltage (V) | Erase voltage (V) | Read voltage (V) | ON/OFF ratio | Switching time |
|---|---|---|---|---|---|---|
| AL/PKEu/ITO | Flash | - 2.0 | + 4.0 | - 1.0 | $10^4$ | 20 μs |
| ITO/PCzOxEu/Al | Flash | - 2.8 | + 4.4 | + 1.0 | $10^5$ | 1.5 μs |
| Al/PF6Eu/ITO | WORM | + 3.0 | N/A | + 1.0 | $10^7$ | N/A |
| Al/PF6Eu/n-Si | WORM | + 2.2 | N/A | + 1.0 | $10^4$ | N/A |
| Al/PF8Eu/ITO | WORM | + 3.0 | N/A | + 1.0 | $10^6$ | ~ 1 μs |
| ITO/PFOxPy/Al | DRAM | - 2.8 | + 3.5 | - 1.0 | $10^6$ | N/A |
| ITO/PCz/Al | WORM | - 1.8 | N/A | - 1.0 | $10^6$ | 1 ms |
| ITO/PVK-$C_{60}$/Al | Flash | - 2.8 | + 3.0 | + 1.0 | $10^5$ | N/A |
| ITO/TPS-PI/Al | WORM | -5.7 | N/A | - 1.0 | $10^5$ | N/A |
| ITO/PP6F-PI/Al | Flash | ~ 4.5 | - 5.0 | + 2.0 | $10^6$ | N/A |
| ITO/TP6F-PI/Al | DRAM | + 3.2 | - 2.1 | + 1.0 | $10^5$ | N/A |
| Al/PVK-AuNP/TaN | Flash | + 3.0 | - 1.7 | + 1.0 | $10^5$ | 1.0 μs |

## 7.6 Summary and Outlooks

The investigation of organic molecule-based information memories has been developing rapidly and many significant results are being reported everyday. However, there are far too many one-time "wanders" in the field of molecular electronics, which don't stand for the test of time because they are based on some "exciting" but artificial observations. So there has to be quantitative measures of the responsivity including thermal/dynamic responses, temperature dependence, molecule/electrode interfaces, junction size effect, molecular structural factors, and theory models for the device performance.

It is very applicable to realize ultra-high density single-molecule scale data storage in the future. We believe that two approaches may have the most potential for promising applications as building blocks in practical nanoscale information storage. One is single-molecule-based memory device with sub-10 nm characteristics built on molecular self-assembly monolayers. Multimode information storage is the other powerful way to make breakthrough in nanoscale data storage. All of these rely on further advances in both experimental and theoretical researches.

Most of the current experiments were conducted at ambient conditions, which were very poor for molecular electronic device characterizations. It may be a good idea to test the target molecule under controlled conditions, especially under ultra-high vacuum and low temperature situations. In this way, we can then carefully examine and extract the correct information of molecular-specific properties from experimental results, which in turn provides a solid foundation for the development and application of molecular-scale memories. Obviously, the researchers firstly have to develop some simple but effective approaches to fabricate reliable, single-molecule-scale and addressable metal-molecule-metal junctions with well-defined electrode/molecule contacts.

## Acknowledgement

JCL acknowledges the supports from Profs. G. J. Szulczewski and S. C. Blackstock, Dr. K.-Y. Kim (Univ. of Alabama); Profs. Luping Yu and H. M. Jaeger (Univ. of Chicago); Prof. Z.Q. Xue (Peking Univ., China) and Prof. P. Mulvaney (Univ. of Melbourne, Australia). Graduate students Han Xiaobo, Han Na and Han Yu give great help in preparing the schematic drawings. Financial support partially comes from NEU young scholar program.

## Table Captions

**Table 1.** Common organic solvents used for molecular monolayer and/or substrate treatments.

**Table 2.** Procedure to make thiolated molecular self-assembly monolayer onto Au surface. Here 1-decanethiol (i.e. $C_{10}SH$) is used as a typical example.

**Table 3.** Procedure for silicon wafer RCA clean.

**Table 4.** Experimental procedure to make organosilane monolayer (e.g. trichlorosilane) on silicon dioxide substrate.

**Table 5.** Silicon wafer clean procedure for thermal oxidation or thin film deposition.

**Table 6.** Functional substituents usually encountered in molecular electronic materials, especially in molecules for monolayer devices.

**Table 7.** Summary and comparison of molecular junction fabrication approaches. [Adapted with permission from Ref. 5. Copyright (2009), World Scientific Publishing Co.]

**Table 8.** Work function of metals used in molecular electronics. Note: the values are just for clean metal electrodes, it may drastically change upon contact with organic layers.

**Table 9.** Comparison of switching mechanisms of molecular memory systems.

**Table 10.** Comparison of the molecular first oxidation potential, thin film morphology and electrical performance of GaIn/50 nm redox film/Ag junctions.

**Table 11.** Comparison of current per molecule ($I$) through saturated sigma-bonded single molecules and monomolecular layers sandwiched between two electrodes measured by a variety of different experimental approaches at two different voltages (0.5 and 0.2 V). In the table, 'Gap' represents the molecular HOMO-LUMO energy gap and 'Length' is calculated molecular length. [Adapted with permission from Ref. 72. Copyright (2003), Wiley-VCH Verlag GmbH & Co. KGaA. Weinheim.]

**Table 12.** Comparison of currents through conjugated pi-bonded single molecules and monolayers sandwiched between two electrodes measured by a variety of different experimental approaches at two different voltages. In this table, 'Gap' represents the molecular HOMO-LUMO energy gap, Length is calculated molecular length. The molecular structure is shown in figure 19. [Adapted with permission from Ref. 72. Copyright (2003), Wiley-VCH Verlag GmbH & Co. KGaA. Weinheim.]

**Table 13.** Comparison of molecular length, apparent height and decay constant β values for a series of molecular materials experimentally measured by using scanning probe microscopy techniques. Molecular structures are given in figure 19 [37,73-76]. [Adapted with permission from Ref. 73. Copyright (2005), American Chemical Society.]

**Table 14.** Effect of molecular conformation on the electronic conductance (unit in $G_0$). The single molecular junction is fabricated by using the method of breaking Au point-contacts in the corresponding molecular solution. [Adapted with permission from Ref. 77. Copyright (2006), Nature Publishing Group.]

**Table 15.** Effect of molecular substituent on the molecular conductance. The ionization potential (IP) and relative conductance are also calculated and given out. [Adapted with permission from Ref. 35. Copyright (2007), American Chemical Society.]

**Table 16.** First-principle molecular electronic structure calculated using Hartree-Fock (HF), density functional theory (DFT) and HF-DFT hybrid methods. [Adapted with permission from Ref. 79. Copyright (2002), The Japan Society of Applied Physics.]

**Table 17.** Summary of electrical properties of organosilane monolayers with different ending groups with junction structure of Al/Silane SAM/Si. The substrate is n type Si (100). The values in parenthesis are calculated ones, while n is the optical indice of the monolayer. [Adapted with permission from Ref. 9. Copyright (1997), Elsevier B.V.]

**Table 18.** Comparison of the fabrication approaches for self-assembly monolayer patterning.

**Table 19.** Silicon wafer preparation for photoresist lithography patterning.

**Table 20.** Experimental procedure for S1813 photoresist photolithography.

**Table 21.** Summary of the device performance evaluation results of polymer memories. [Adapted with permission from Ref. 97. Copyright (2007), Elsevier B.V. ]

## Figure Captions

**Figure 1.** Schematic correlation between the main researching contents of molecular electronics.

**Figure 2.** Schematic vacuum thermal vapor deposition system with in-situ vary-temperature current-voltage characterizations. Here, 1– Power, 2–evaporation crucible heating controllers, 3– to ultra-high vacuum pumps, 4– sample, 5– temperature sensor, 6– shutter, 7- liquid nitrogen feedthrough, 8– current/voltage measurement, 9– temperature detector, 10– substrate heating controller, 11– film thickness monitor, 12– LED light, 13– data processing board, 14–computer, 15– thermal insulator, 16– substrate holder, 17– electrical bridging connector.

**Figure 3.** Schematic self-assembly monolayer of (a) 1-decanethiol and (b) 1-nonanedithiol sandwiched between Au electrodes.

**Figure 4.** Schematic self-assembly process for trichlorosilane monolayer formation on silicon dioxide surface. A very thin water layer is a necessary factor in this chemical process. [Adapted with permission from Ref. 9. Copyright (1997), Elsevier B.V.]

**Figure 5.** Schematic process for LB monolayer fabrication. (a) droplet of molecular solution is spread on the water surface in the trough. (b) The molecules are compressed by slowly moving the barrier bar. (c) The pre-dipped substrate is pull out of the water with the molecules transferred onto its surface. (d) The as-fabricated sample with a single LB layer.

**Figure 6.** Schematic diagram of vacuum spray deposition system. 1–Polymer solution, 2–Compress pump, 3–Valve, 4–Flow meter, 5–Nozzle, 6–Pressure gauge, 7–Observation window, 8–Vacuum chamber, 9–Film thickness monitor, 10–Substrate heating electrical feedthrough, 11–Substrate holder, 12–Sample, 13–To high vacuum pump system and 14–Zoom of the spray.

**Figure 7.** Several ring compounds containing nitrogen, oxygen and sulfur incorporated in heterocyclic system, which are used as building units in the design of molecular electronic materials. The molecules are named as (a) Benzene, (b) Pyridine, (c) Pyrimidine, (d) Phenol, (e) Furan, (f) Thiophene, (g) Pyrrole, (h) Quinoline, (i) Purine and (j) Indole.

**Figure 8.** Schematic structure of a junction with one molecular wire contacted with left (L) and right (R) electrodes. The left and right electrode/molecule interfaces can drastically affect the junction's charge transport and performance. The molecular functional units are labeled from 1 to N.

**Figure 9.** Chemical structure of a series of redox arylamine molecular materials. The first Group includes molecules **a, b** and **c**, the second group are molecules **d–i**, while the left redox arylamine molecules are classified as the third group. Here, molecule **i** is widely known as TPD, a hole transport materials used in organic light emitting diodes.

**Figure 10.** Representative I-V curves for six typical materials. It is clear that the turn-on voltage of different molecular thin films are drastically different although the active layer was kept in the same thickness. Inset shows the schematic device structure and measuring system used [13].

**Figure 11.** Tapping mode AFM images for the thin films of figure 9 molecules **a**, **h**, **n**, **k** and **l** (from top to down), respectively. The films are 50-nm-thick vapor deposited on Ag electrodes. Low molecular weight materials with symmetry structures tend to form rough and loosely packed films, while moderate molecular weight and dendrimeric materials usually show smooth and close-packed characters [13].

**Figure 12.** Effect of anode metal work function on the device turn-on voltage for molecule **k**.

**Figure 13.** Schematic drawings show the charge transfer (CT) interfaces between redox molecular thin films with electron acceptor group (–CN) on substrates: (a) without buffer layer, (b) with buffer layer, (c) buffer layer on SAM and (d) acceptor film on SAM. Buffer layer means the molecular thin films made from molecules without electron acceptor group. The molecular structure of the two kinds of materials is also shown.

**Figure 14.** Chemical structure of the organic materials studied for understanding the electronic structure and electrical properties of interfaces between metals and pi-conjugated molecular thin films [46].

**Figure 15.** Comparison of the molecular ionization energy values of organic materials measured by ultraviolet photoemission spectroscopy (UPS). Here, IE is defined as the energy difference between the leading edge of the HOMO and the vacuum level obtained from the photoemission cutoff. The chemical structure of the molecules is given in a figure 14. [Adapted with permission from Ref. 46. Copyright (2003), Wiley Periodicals, Inc.]

**Figure 16.** Comparison between metal work function, IE and electron affinity (EA) (i.e., HOMO and LUMO) positions of various molecular materials. The zero is defined as the vacuum level. The IE and EA are determined by UPS and inverse photoemission spectroscopy. [Adapted with permission from Ref. 46. Copyright (2003), Wiley Periodicals, Inc.]

**Figure 17.** Schematic experimental setups for charge storage or electrical switching on redox dendrimeric thin films or monolayers. (a) Molecular junction of Ag/TPD/4AAPD/TPD/Ag [7], (b) Au/Molecule in $SiO_2$/Au [47], (c) Au-tip/Molecules in PMMA/$SiO_2$/Si [49], (d) Au/SAM/electrolyte [50] and (e) STM tip/Molecule/Au [53-55].

**Figure 18.** Chemical structure of the molecules listed in table 12.

**Figure 19.** Chemical structures for the molecules studied by using STM shown in table 13.

**Figure 20.** Chemical structure of the molecules listed in table 14.

**Figure 21.** Molecular structures of a series of conjugated molecules calculated.

**Figure 22.** Organosilane monolayer can be used as surface anchor to direct the growth of a variety of conjugated parts on the substrate (from left to right: phenyl, pyrene and retinal ended monolayer). This technique will be useful in the bottom-up approach of molecular memory system. [Adapted with permission from Ref. 9. Copyright (1997), Elsevier B.V.]

**Figure 23.** Organosilane monolayers can be used to modify the substrate surface and direct the assembly of metal nanoparticles. (a) $–COO^-$ terminated SAM is selectively formed on gold surfaces and (b) $–NH_3^+$ terminated molecules are selectively formed onto silicon dioxide surfaces [11]. (c) Silver nanoparticle is selectively attached to the $–NH_2$ terminated aminosilane surfaces [12].

**Figure 24.** Schematic structure showing a conjugated dithiol molecular wire inserted within 1-dodecanethiol matrix.

**Figure 25.** Schematic process for patterning SAM on substrate: (a) Stamp-printing, (b) selective growth on pre-patterned substrate (e.g. lithography pre-defined Au regions), (c) AFM tip-based dip-pen writing and (d) STM tip-based lithography.

**Figure 26.** Chemical structure of molecular wires studied (top). Effect of molecular structure on the current switching ratio between light and dark conditions (bottom). It is clear that the junction with the conjugated molecules can yield a higher switching ratio than that without [6].

**Figure 27.** *I–V* temperature dependence of a 1-decanethiol junction (a) in the dark and (b) under illumination of florescent light. [Reprinted with permission from Ref. 6. Copyright (2009), Elsevier B.V.]

**Figure 28.** Molecular structure of the polymers with electrical switching properties [97].